\documentstyle[epsf,emulateapj,apjfonts]{article}

\slugcomment{The Astrophysical Journal, in press}
\lefthead{SARAZIN, WISE, \& MARKEVITCH}
\righthead{X-RAY SPECTRUM OF A2029}

\begin{document}
 
\title{X-Ray Spectral Properties of the Cluster Abell 2029}

\author{Craig L. Sarazin}
\affil{Department of Astronomy, University of Virginia, \\
P.O. Box 3818, Charlottesville, VA 22903-0818; \\
cls7i@virginia.edu,}

\author{Michael W. Wise}
\affil{Center for Space Research, Massachusetts Institute of Technology, \\
Bldg.\ 37-644, Cambridge, MA 02139; wise@space.mit.edu,}

\and
 
\author{Maxim L. Markevitch}
\affil{Harvard-Smithsonian Center for Astrophysics, \\
60 Garden Street, Cambridge, MA 02138; \\
maxim@head-cfa.harvard.edu}

\begin{abstract}
We have analyzed {\it ASCA} and {\it ROSAT} PSPC spectra and images
of the galaxy cluster Abell 2029.
The {\it ASCA} spectra of the cluster indicate that the gas
temperature declines with radius.
The PSPC image shows that the cluster is very regular and smooth.
Also, there is no significant evidence for any irregularities in the
temperature distribution in the cluster, as would be produced by a
subcluster merger.
These results suggest that A2029 is a relaxed cluster, and that the
gas is in hydrostatic equilibrium.
We use the assumption of equilibrium to determine the gravitational mass
of the cluster as a function of radius.
At a radius of $16'$ ($1.92 h_{50}^{-1}$ Mpc;
$H_o = 50 h_{50}$ km s$^{-1}$ Mpc$^{-1}$),
the gravitational mass is
$M_{tot} = ( 9.42 \pm 4.22 ) \times 10^{14} h_{50}^{-1} \, M_\odot$,
while the mass of gas is
$M_{gas} = ( 2.52 \pm 0.77 ) \times 10^{14} h_{50}^{-5/2} \, M_\odot$.
The gas fraction is found to increase with radius;
within a spherical radius of $16'$, the fraction is
$M_{gas}/M_{tot} = ( 0.26 \pm 0.14 ) h_{50}^{-3/2} $.

The iron abundance in the gas is found to be $0.40 \pm 0.04$ of solar.
There is no significant evidence for any variation in the abundance
with position in the cluster.
The global X-ray spectra, central X-ray spectra, and {\it ROSAT}
surface brightness all require a cooling flow at the cluster center.
The global X-ray spectrum implies that the total cooling rate is
$363^{+79}_{-96} h_{50}^{-2} \, M_\odot$ yr$^{-1}$.

The global X-ray spectra are consistent with the Galactic
value for the soft X-ray absorption toward the cluster.
The {\it ROSAT} PSPC spectra of the central regions of the
cluster are inconsistent with the large value of foreground
excess absorption found by White et al.\ (1991)
based on the {\it Einstein} SSS spectrum.
The upper limit on excess foreground absorption is
$7.3 \times 10^{19}$ cm$^{-2}$.
However, the spectra do not rule a significant amount of intrinsic
absorbing gas located within the cooling flow region.
\end{abstract}


\section{INTRODUCTION} \label{sec:intro}

Abell 2029 (A2029) is a richness class II, Bautz-Morgan class I cluster of
galaxies at a redshift of $z = 0.0767$
(Abell, Corwin, \& Olowin 1989).
With an X-ray flux of
$F_X$ (2--10 keV) $= 7.52 \times 10^{-11}$ ergs cm$^{-2}$ s$^{-1}$
(David et al.\ 1993), it is one of the brighter X-ray clusters.
This flux corresponds to an X-ray luminosity of
$L_X$ (2--10 keV) $= 2.07 \times 10^{45} h_{50}^{-2}$ ergs s$^{-1}$, and a
bolometric X-ray luminosity of
$L_X (bol) = 4.26 \times 10^{45} h_{50}^{-2} $ ergs s$^{-1}$
($H_o = 50 h_{50}$ km s$^{-1}$ Mpc$^{-1}$).

A2029 is one of the most regular rich clusters known, particularly in its
X-ray properties
(Slezak, Durret, \& Gerbal 1994;
David, Jones, \& Forman 1995;
Buote \& Tsai 1996).
As a result, it seems likely that the intracluster gas in A2029 is in
hydrostatic equilibrium out to a large radius.
Thus, A2029 is an ideal candidate for the determination of the total
gravitational mass of the cluster at large radii using the assumption
of hydrostatic equilibrium.
This requires that the gas temperature and density be determined as a
function of radius the cluster. 
Previous observations using the $Einstein$ Monitor Proportional Counter
(MPC) found an average gas temperature of
$k T = 7.8^{+1.4}_{-1.0}$ keV
(David et al.\ 1993).
However, this observation does not provide any information on the
distribution of temperatures.
David et al.\ (1995) derived the mass of A2029 from
{\it ROSAT} Position Sensitive Proportional Counter (PSPC) observations
assuming that the gas was isothermal.

This paper presents new {\it ASCA} X-ray spectra of A2029.
We use these spectra to constrain the temperature variation in the
cluster.
We also analyze archival {\it ROSAT} PSPC observations of the X-ray
image and spectrum of A2029.
We use the {\it ASCA} derived temperature gradients and {\it ROSAT} derived
X-ray surface brightness to determine the variation of the total
gravitational mass and gas mass with radius in the cluster.
As a test of the hydrostatic assumption,
we search for asymmetries in the the temperature distribution which
would be indicative of recent dynamical activity, such as a subcluster
merger.
We also constrain any abundance gradients in the intracluster gas
in the cluster.
The use of {\it ASCA} spectra to determine the spatial variation of
the spectral properties of the X-ray emitting gas in A2029 requires that
the effects of the energy-dependent Point Spread Function (PSF) be
corrected.
This is particularly difficult when the X-ray surface brightness is
strongly peaked to the cluster center, as is true in A2029.

A2029 contains a very strong cooling flow, which is centered on the
large central cD galaxy (UGC 9752 = IC 1101).
The presence of a cooling flow has been established based both on
analyses of the X-ray surface brightness profile
(Arnaud 1989; Sarazin, O'Connell, \& McNamara 1992),
and on analyses of the X-ray spectrum of the central cluster region
(White et al.\ 1991).
The {\it Einstein} Imaging Proportional Counter (IPC) observations
of the surface brightness gave a cooling rate of
$\dot{M}_{cool} = 366 h_{50}^{-2} \, M_\odot$ yr$^{-1}$ within a
radius of $1\farcm9 = 226 h_{50}^{-1}$ kpc.
(Arnaud 1989),
while the {\it ROSAT} High Resolution Imager (HRI) gave
$\dot{M}_{cool} = 370 h_{50}^{-2} \, M_\odot$ yr$^{-1}$ within
$228 h_{50}^{-1}$ kpc
(Sarazin et al.\ 1992).
The {\it Einstein} Solid State Spectrometer (SSS) spectrum gave
a cooling rate of
$\dot{M}_{cool} = 513^{+304}_{-247} h_{50}^{-2} \, M_\odot$ yr$^{-1}$
within its aperture of 3$'$
(White et al.\ 1991).
However, A2029 is nearly unique among strong, well-studied cooling flows
in that it shows no evidence for optical emission lines or blue stellar
continuum color within its central regions
(e.g., McNamara \& O'Connell 1989).

White et al.\ (1991) reanalyzed the $Einstein$ SSS spectra of the central
regions (3 arcmin radius) of a number of cooling flow clusters,
and found evidence for large amounts of soft X-ray absorption,
in excess of that expected from the interstellar gas in our Galaxy.
In A2029,
White et al.\ found an excess absorbing column of
$\Delta N_H = 15 \times 10^{20}$ cm$^{-2}$,
if the excess absorber was modeled as foreground Galactic material.
Because the Galactic hydrogen column towards A2029 is relatively small,
$N_H = 3.07 \times 10^{20}$ cm$^{-2}$ (Stark et al.\ 1992), this
excess absorption should have a profound effect on the soft X-ray
spectrum of the central regions of A2029.

We extract the central {\it ASCA} and {\it ROSAT} X-ray
spectra of A2029 to study the spectral properties of the cooling flow
gas.
In particularly, we use the soft X-ray spectra to constrain
the amount of absorbing material in excess of that expected from
our Galaxy.

In \S~\ref{sec:obs}, we describe the {\it ASCA} and {\it ROSAT}
observations and the treatment of the data.
The {\it ROSAT} PSPC image  of the cluster field is extracted in
\S~\ref{sec:image};
it is needed to analyze the spectral data and to determine the
X-ray surface brightness profile.
The global X-ray spectrum is analyzed using {\it ASCA} and {\it ROSAT} data
in \S~\ref{sec:global}.
In \S~\ref{sec:spatial}, we discuss the spatial variation of the X-ray
spectrum of A2029.
Here, we determine the radial temperature and abundance gradients, and
search for asymmetries in the temperature distribution.
The central cooling flow spectrum is extracted in \S~\ref{sec:cflow}.
Using the X-ray temperature and surface brightness profiles, we determine
the distribution of the total gravitational mass and gas mass in
\S~\ref{sec:masses}.
Finally, our conclusions are summarized in \S~\ref{sec:conclude}.

Unless otherwise noted, all uncertainties quoted in this paper
are 90\% confidence regions.
The abundances given are all relative to the ``cosmic'' values 
in Anders \& Grevesse (1989).
All distance-dependent values in this paper assume
$H_o = 50$ km s$^{-1}$ Mpc$^{-1}$ and $q_o = 0.0$;
we have tried to indicate the distance dependence by parameterizing
the Hubble constant as $H_o = 50 h_{50}$ km s$^{-1}$ Mpc$^{-1}$.

\section{X-RAY OBSERVATIONS} \label{sec:obs}

\subsection{{\it ASCA} Data} \label{sec:obs_asca}

The cluster was observed with {\it ASCA} on
1994 February 19--20.
The Solid-state Imaging Spectrometer (SIS) detectors operated with two
chips per detector being active
(S0C0,1 and S1C2,3), and the observation included both bright and
faint mode SIS data.
There were a number of problems with the attitude control
during a portion of this observation, which resulted in
the original standard attitude file being incorrect.
For the first 1180 seconds of the integration, the attitude was about
20 arcmin off.
We discarded this portion of data.
The attitude settled into an accurate and steady value for the remainder
of the observation.
The Gas Imaging Spectrometer detector 3 (GIS3) data was taken during the
period when the events were recorded with a bit error.
This error was corrected, although this data has fewer spectral
channels as a result of this.
The data were screened using relatively strict versions of the standard
cleaning criteria.
When these screening operations were complete, the exposures were
29,612, 29,710, 34,365 and 34,162 seconds for the SIS0, SIS1, GIS2, and 
GIS3, respectively.

The SIS and GIS spectra were corrected for background using blank sky
observations which were screened in the same manner as the data.
For the GIS, the background was constructed by weighting blank sky
observations with differing values of the cut-off rigidity ($COR$)
by the actual distribution of $COR$ values during the screened exposure
time.
For the SIS, a long observation with the same minimum cut-off rigidity
($COR > 8$) was used.
For the analysis of the spatial variation of the X-ray spectrum
(\S~\ref{sec:spatial}), we added a systematic uncertainty
of 20\% in the background to the other uncertainties in deriving confidence
intervals for parameters.

\subsection{{\it ROSAT} PSPC Data} \label{sec:pspc_data}

A2029 was observed
with the {\it ROSAT} Position Sensitive Proportional Counter (PSPC)
for 3,236 seconds on 1992 January 24
(observation wp800161; U. Briel, P.I.)
and for 12,550 seconds on 1992 August 10-11
(observation rp800249; C. Jones, P.I.).
The first observation was discussed previously in
Slezak et al.\ (1994), while the second was presented in
David et al.\ (1995).
The two observations were combined, and were aligned using the positions
of two bright point sources.
The PSPC data was screened for periods of high background based on
a Master Veto Rate $>$ 170
(Plucinsky et al.\ 1993),
for other times of high background based on the light curve of
counts in the background region used for the extraction of the
spectrum (see below),
for periods of 15 seconds after switching on the high voltage,
and for periods with an uncertain aspect solution.
The analysis of the PSPC was done with a combination of the IRAF/PROS
software, the XSPEC/XANADU software, the XSELECT program, and Snowden's
SXRB package for the analysis of diffuse X-ray emission
(Snowden 1995).
The screened data contained a total exposure of 12,905 seconds
and the average Master Veto Rate was 87.05.
The data were corrected for particle background, scattered
solar X-ray background, the long term enhancements (LTE) in this
background, and after-pulses.
With these backgrounds removed, the image was corrected for
exposure and vignetting.

The {\it ROSAT} PSPC spectra and X-ray surface brightness profiles were
corrected for background using
four regions at radii from 24 to 44 arcmin which were free of strong
sources or obstruction by the instrument ribs.
Because of calibration uncertainties at the softest and hardest
energies, we restricted the PSPC spectra to energies of 0.2--2.2 keV
(channels 6-32).

\subsection{Spectral Modeling} \label{sec:spect_model}

We fit the observed spectra with thermal emission models for
a low density plasma in collisional ionization equilibrium.
Where possible, we used models based on both the Raymond-Smith code 
(Raymond \& Smith 1977) and the MEKAL code
(e.g., Kaastra et al.\ 1996).
The MEKAL code includes a better treatment of the iron L lines
near 1 keV, based on measurement and calculations of the
atomic parameters by Liedahl and his colleagues
(e.g., Liedahl et al.\ 1995).
We did not fit MEKAL models during the spatial analysis of the spectra in
\S~\ref{sec:spatial}, as the MEKAL spectra were not available in a form
which could be used with the programs employed to correct for the
energy-dependent Point-Spread-Function (PSF).
However, at higher temperatures, we found that the Raymond-Smith
and MEKAL spectra gave almost exactly the same results.
Larger differences are found between the Raymond-Smith and MEKAL
spectra at lower temperatures, and these become more important when
the spectrum includes the cooling flow regions at the center of the
spectra.
However, we still found that the spectral parameters from fitting
Raymond-Smith and MEKAL models were in good agreement.
Because the MEKAL model generally gave slightly better fits, we report
the results using this model in \S\S~\ref{sec:global} and 
\ref{sec:cflow}.

We considered two simple models for the thermal distribution of
the gas in each region where the spectrum was extracted.
The first model assumed thermal emission at a single temperature;
we refer to this as a ``single temperature'' model.
In Table~\ref{tab:spectra}, these models labeled as ``1T.''
This model is characterized by two spectral parameters,
the temperature of the gas and the ratio of the abundances
in the gas to the solar values, and by the overall
emission integral of the gas which normalizes the spectra.

As discussed above (\S~\ref{sec:intro}), there is considerable
previous evidence from X-ray spectra and images for a cooling flow
at the center of A2029.
Thus, we also considered a simple spectral model for ambient gas and
a cooling flow.
The gas in the cooling flow was assumed to cool from the temperature
of the ambient gas to very low temperatures due to its own radiation.
The abundances in the cooling flow gas were taken to be those in the
ambient gas, as expected if the cooling gas originated as intracluster
gas.
The cooling flow gas was assumed to cool at constant pressure;
more detailed spectral models have indicated that this is a
reasonably accurate approximation (Wise \& Sarazin 1993).
The X-ray spectrum of this cooling flow component is given by
\begin{equation} \label{eq:cflow}
L_E = \frac52 \, \dot{M}_{cool} \, \frac{k}{\mu m_p}
\, \int_o^{T} \frac{\Lambda_E ( T' )}{\Lambda_{tot} ( T' )}
\, d T' \, ,
\end{equation}
where $E = h \nu$ is the photon energy, $L_E \, dE$ is the luminosity
emitted in the photon energy range $E \rightarrow E + dE$,
$\mu \approx 0.61$ is the mean mass per particle in terms of the proton
mass $m_p$,
$T$ is the ambient gas temperature in the cluster from which the
gas is assumed to cool,
$T'$ is its current temperature,
the emissivity of the gas per unit volume and photon energy is
$\rho^2 \Lambda_E (T)$,
$\rho$ is the mass density of the gas, and the total emissivity
(cooling rate) of the gas is $\rho^2 \Lambda_{tot} ( T )$
(e.g., Wise \& Sarazin 1993).
We will refer to this model as the ``single temperature plus
cooling flow model.''
In Table~\ref{tab:spectra}, these models labeled as ``CF.''
This model is characterized by a single additional parameter,
the cooling rate $\dot{M}_{cool}$.

The global cluster X-ray spectrum (\ref{sec:global}) and the
spectrum of the inner cooling flow region (\S~\ref{sec:cflow})
were fit to these models using XSPEC version 9.
Because this version of XSPEC did not contain a model for
the cooling flow spectrum using the MEKAL emission code, we constructed
such a model.
The analysis of the spatial variation of the spectrum, correcting for
the effects of the energy-dependent PSF of {\it ASCA}, was done with
the program of Markevitch et al.\ (1996).

\section{{\it ROSAT} PSPC IMAGE} \label{sec:image}

The {\it ROSAT} PSPC data in the hard band of 0.5 to 2.0 keV
(PI channels 52 to 201) were used to construct an image of the region
of the {\it ASCA} observation.
This hard band is useful because it reduces the background in the image
without much loss of counts from the hot intracluster gas.
This image was corrected for particle and solar X-ray background, exposure, and
vignetting as discussed above.
The resulting image was adaptively smoothed with a variable smoothing
kernel so that each smoothing beam had a signal to noise ratio of
at least five
(Huang \& Sarazin 1996).
Contours of the X-ray surface brightness for the central portion of the
resulting image are shown in
Figure~\ref{fig:pspc_image}.

\centerline{\null}
\vskip2.7truein
\includegraphics{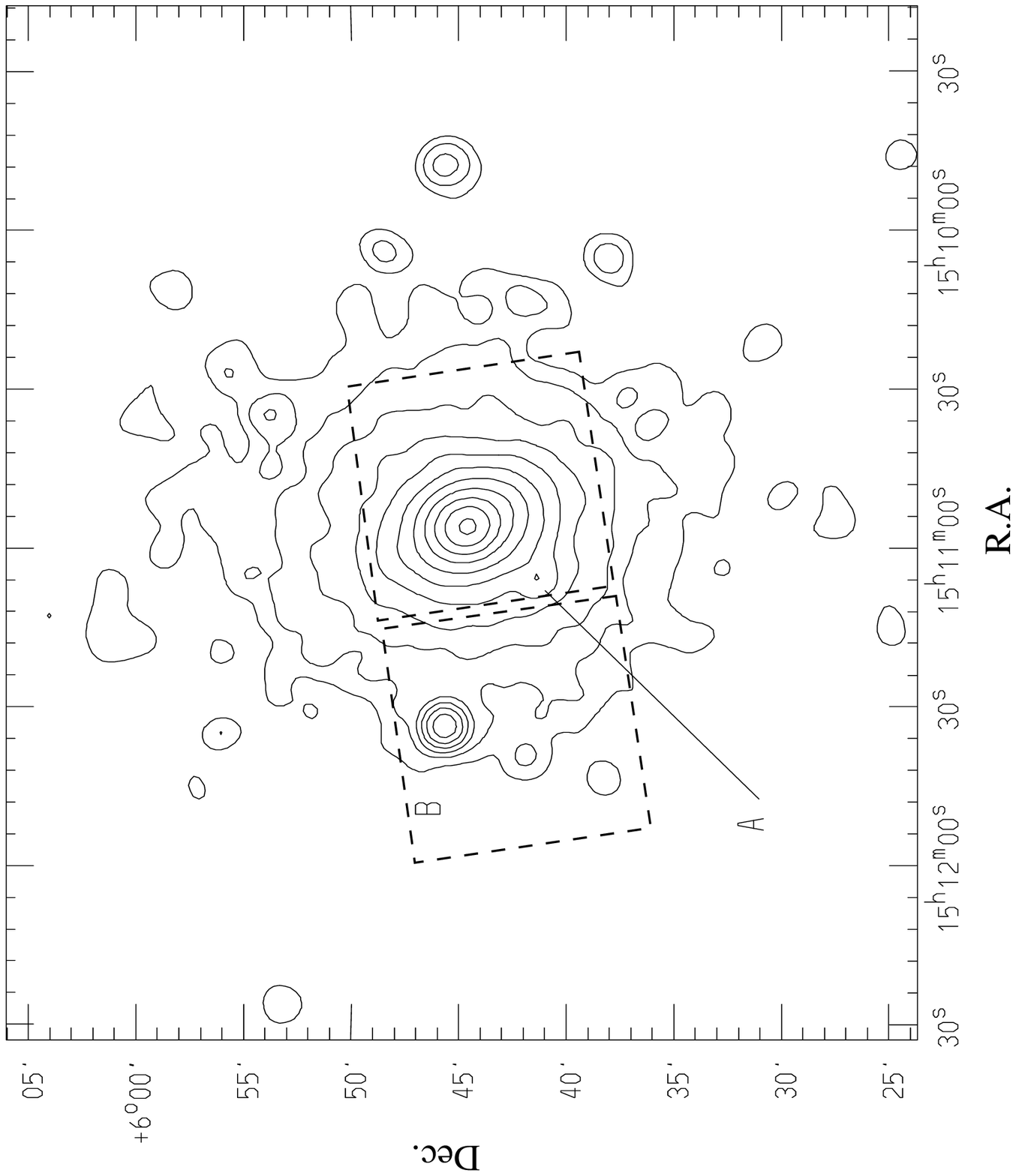}
\figcaption
{Contours of the {\it ROSAT} PSPC image in the hard band
(0.5 to 2.0 keV) of the A2029 cluster field,
corrected for particle and solar X-ray background
and for exposure.
The image was adaptively smoothed to a signal to noise ratio of
five per smoothing beam.
The contours are logarithmically spaced with four contours per
dex,
and the lowest contour corresponds to 0.001 counts s$^{-1}$ arcmin$^{-2}$.
The coordinates are J2000.
Two unrelated point sources located approximately 4 and 9 arcmin
from the cluster center are labeled as ``A'' and ``B,'' respectively.
The two dashed squares give the approximate fields of view of the two
chips of the SIS0 detector on {\it ASCA}.
\label{fig:pspc_image}}

\vskip0.2truein

The {\it ROSAT} PSPC image of A2029 contains a large number of
presumably unrelated point sources.
There are two moderately strong point sources located
approximately 4 arcmin (Src.\ A) and 9 arcmin (Src.\ B) from
the center of the cluster.
Src.\ A is known to be a background active galaxy nucleus (AGN;
Bregman, private communication).
Src.\ B has not been identified to our knowledge.
These sources are labeled on Figure~\ref{fig:pspc_image}, and
are discussed further in \S~\ref{sec:spatial},
where they are included in the determination of the variation in the
X-ray spectrum of A2029.

When these sources are removed, the contours of the X-ray surface
brightness of A2029 in Figure~\ref{fig:pspc_image}
appear elliptical and quite regular.
This agrees with the previous conclusions reached
by Slezak et al.\ (1994) using wavelet techniques applied to the
shorter {\it ROSAT} PSPC exposure, and
by Buote \& Tsai (1996) using the power ratio technique applied to the
longer {\it ROSAT} PSPC exposure.
There is general agreement that A2029 is regular and that it lacks
any strong substructure.
This is also consistent with the tendency for clusters
with large cooling flows to be regular (e.g., Buote \& Tsai 1996).
The regularity of A2029 makes it a particularly attractive candidate
for applying the hydrostatic method to determine the mass distribution.

At the faintest levels, the {\it ROSAT} PSPC observation shows
X-ray emission from A2029 extending out to a radius of approximately
20 arcmin from the cluster center.
In \S~\ref{sec:spatial}, we will analyze the X-ray spectrum of
A2029 out to a radius of 16 arcmin.
All of this region is included in the fields of view of both
of the GIS detectors.
However, the SIS detectors only cover a fraction of the cluster
emission.
In Figure~\ref{fig:pspc_image},
the dashed squares give the approximate fields of view of the two
chips used on the SIS0 detector (S0C0 and S0C1).
The corresponding regions for the SIS1 detector chips
(S1C2 and S1C3) are similar and have about the same orientation,
but are displaced about 1 arcmin to the west.

\section{GLOBAL X-RAY SPECTRUM} \label{sec:global}

The {\it ASCA} and {\it ROSAT} observations were used to determine
the integrated spectrum of the cluster.
For this purpose, we determined the spectrum within 16$'$
($1.92 h_{50}^{-1}$ Mpc) of the center of the cluster.
This region is shown as the outermost dashed circle in
Figure~\ref{fig:pspc_slices}.
This region extends nearly to the bright ring of enhanced background
near the edge of each of the {\it ASCA} GIS detectors, and lies well within
the ribs on {\it ROSAT} PSPC detector.
Since the {\it ASCA} SIS spectra cover only a portion of this region
(Fig.~\ref{fig:pspc_image}), we do not include the SIS data in this
determination of the global spectrum.
Later, we will find that the cluster has significant spatial variations
in its X-ray spectrum, due to a cooling flow at the cluster center
(\S~\ref{sec:cflow}) and due to a radial temperature gradient
(\S~\ref{sec:spatial_t}).
Thus, it would be difficult to directly compare the SIS spectra with
those from the GIS and PSPC, which cover the same larger region of the cluster.
The spectra were binned to insure that there were at least 20 counts
per spectral channel, so that the $\chi^2$ test gave reliable results.
We excluded appropriately sized regions around each of the two
point sources labeled A and B in Figure~\ref{fig:pspc_image}.

The results of fitting ``single temperature'' and ``single temperature plus
cooling flow'' models to the observed spectral are summarized
in Table~\ref{tab:spectra}.
The first column enumerates the rows for reference in the text.
The second column indicates the region of A2029 being considered, and
the third column lists the instrument(s) used.
The fourth column gives the model being fit to the data.
Columns 5--8 give the galactic hydrogen absorbing column $N_H$,
the ambient gas temperature $kT$, the heavy element abundance, and
the cooling rate (the latter is only given for the CF models).
Column 9 gives the resulting $\chi^2$ and number of degrees
of freedom (d.o.f.) for the best-fit model.
The final column gives the probability that the value of $\chi^2$
would be this large if the data were drawn from the best-fit model
with only statistical uncertainties.
Because it is likely that there are systematic errors in the responses
of the instruments, these numbers may not be accurate.
The uncertainties in the parameters give the 90\% confidence region.

Because of their nearly disjoint spectral bands,
we find that the {\it ROSAT} PSPC (0.2--2 keV) and {\it ASCA} GIS's
(0.5--11 keV) provide nearly orthogonal but complementary information
of the spectrum of A2029.
We first discuss the results from {\it ROSAT} and {\it ASCA} separately,
and then jointly fit all of the spectra.

\begin{table*}[tbh]
\small
\caption[ASCA and ROSAT X-Ray Spectral Fits]{}
\label{tab:spectra}
\begin{center}
\begin{tabular}{llcccccccc}
\multicolumn{10}{c}{\sc ASCA and ROSAT X-Ray Spectral Fits} \cr
\tableline
\tableline
Row&Region&Instr.\tablenotemark{a}&Model\tablenotemark{b}&
$N_H$&$kT$&Abund.&$\dot{M}_{cool} h_{50}^2$&$\chi^2$/d.o.f.&Prob.\cr
&&&&($10^{20}$ cm$^{-2}$)&(keV)&($\odot$)&($M_\odot$ yr$^{-1}$)
&&(\%)\cr
\tableline
1&Whole Cluster&PSPC&1T&
$3.21^{+0.22}_{-0.23}$&$6.13^{+1.13}_{-0.82}$&$0.99^{+1.26}_{-0.56}$&
---&$34.05/23 = 1.48$&\phn6.44\cr
2&$(r < 16')$&PSPC&CF&$2.81^{+0.20}_{-0.16}$&$>$6.23&$>$0.34&
$377^{+77}_{-72}$&$24.47/22 = 1.11$&32.31\cr
3&&GIS&1T&$<$4.43&$8.69^{+0.28}_{-0.30}$&$0.37^{+0.04}_{-0.03}$&
---&$788.0/727 = 1.08$&\phn5.77\cr
4&&GIS&CF&$<$6.04&$9.54^{+1.52}_{-1.00}$&$0.40 \pm 0.04$&
$<$609&$786.0/726 = 1.08$&\phn6.05\cr
5&&PSPC,GIS&1T&$3.21 \pm 0.11$&$8.06 \pm 0.21$&$0.38 \pm 0.03$&
---&$860.1/753 = 1.14$&\phn0.64\cr
6&&PSPC,GIS&CF&$3.29 \pm 0.12$&$9.35^{+0.55}_{-0.45}$&$0.40 \pm 0.04$&
$363^{+79}_{-96}$&$821.9/752 = 1.09$\phn3.87\cr
&&&&&&&&\cr
7&$0' \le r \le 1.5'$&ASCA&CF&3.07&$9.42^{+1.27}_{-1.55}$&0.40&
$216^{+90}_{-120}$&$172.61/197 = 0.88$&\cr
8&$1.5' \le r \le 5'$&ASCA&1T&3.07&$8.68^{+2.04}_{-1.28}$&0.40&&---&\cr
9&$5' \le r \le 10'$&ASCA&1T&3.07&$6.35^{+2.37}_{-1.44}$&0.40&&---&\cr
10&$10' \le r \le 16'$&ASCA&1T&3.07&$6.29^{+3.82}_{-2.01}$&0.40&&---&\cr
&&&&&&&\cr
11&$-20^\circ \le PA \le 70^\circ$&ASCA&1T&3.07&$8.85^{+4.33}_{-3.15}$&0.40&&
$173.70/212 = 0.82$&\cr
12&$70^\circ \le PA \le 160^\circ$&ASCA&1T&3.07&$5.65^{+4.72}_{-2.76}$&0.40&
&---&\cr
13&$160^\circ \le PA \le 250^\circ$&ASCA&1T&3.07&$7.28^{+4.10}_{-2.70}$&0.40&&
---&\cr
14&$250^\circ \le PA \le 340^\circ$&ASCA&1T&3.07&$12.09^{+10.56}_{-6.46}$&0.40&&
---&\cr
&&&&&&&&\cr
15&Center&GIS&1T&$3.75^{+1.41}_{-1.39}$&$7.60^{+0.33}_{-0.31}$&
$0.38\pm0.05$&---&$624.1/557 = 1.12$&\phn2.53\cr
16&$(r < 3')$&SIS&1T&$6.64\pm0.38$&$6.55^{+0.18}_{-0.17}$&$0.41\pm0.04$&---&
$1018.0/816 = 1.25$&$<$0.01\cr
17&&PSPC&1T&$3.25^{+0.15}_{-0.13}$&$4.75^{+0.73}_{-0.58}$&
$0.37^{+0.31}_{-0.22}$&---&$27.1/23 = 1.181$&25.17\cr
18&&GIS&CF&$11.45^{+1.02}_{-4.77}$&$9.61^{+4.58}_{-1.73}$&$0.38\pm0.05$&
$429^{+129}_{-234}$&$618.2/556 = 1.11$&\phn3.44\cr
19&&SIS&CF&$7.62^{+0.50}_{-0.49}$&$7.61^{+0.45}_{-0.38}$&$0.41\pm0.04$&
$200^{+51}_{-50}$&$975.30/815 = 1.20$&$<$0.01\cr
20&&PSPC&CF&$2.89\pm0.23$&$>$6.36&$>$0.30&$329^{+86}_{-166}$&
$20.1/22 = 0.91$&57.68\cr
21&&All&CF&$3.29\pm 0.09$&$7.58^{+0.28}_{-0.16}$&$0.43\pm0.03$&$<$32&
$1959.5/1401 = 1.40$&$<$0.01\cr
&&&&&&&&\cr
\tableline
\end{tabular}
\end{center}
\tablenotetext{a}{The instruments are:
PSPC: {\it ROSAT} PSPC;
GIS: {\it ASCA} GIS2 and GIS3;
SIS: {\it ASCA} SIS0 and SIS1;
ASCA: {\it ASCA} GIS2, GIS3, SIS0, \& SIS1;
All: PSPC, SIS, \& GIS.}
\tablenotetext{b}{1T is the single temperature model, while
CF is the single temperature plus cooling flow model.}
\end{table*}

\subsection{{\it ROSAT} PSPC Global X-ray Spectrum} \label{sec:global_pspc}

The {\it ROSAT} PSPC spectrum of the inner 16$'$ of A2029 was extracted
and corrected for background as discussed above and in
\S~\ref{sec:pspc_data}.
First, a single temperature model was fit to this spectrum;
the results are shown in the first row of Table~\ref{tab:spectra}.
The fit was not acceptable, with a $\chi^2 = 34.05$ for 23 d.o.f.
This unacceptable fit gave a well-defined value for the absorbing
column, but fairly uncertain values for the temperature and abundance
of the gas.
For the high ambient gas temperature in A2029, the low energy band of
{\it ROSAT} makes it relatively insensitive to changes in the
temperature.
Moreover, at these high temperatures, the strongest spectral feature
due to heavy elements is the 7 keV iron K line complex, which is
beyond the band of {\it ROSAT}.
On the other hand, the low energy band of the PSPC does make it
very sensitive to the Galactic absorbing column.
The absorbing column required by the PSPC spectrum is consistent
with the Galactic value of $N_H = 3.07 \times 10^{20}$ cm$^{-2}$
(Stark et al.\ 1992).
However, the temperature is lower than that given previously by the
$Einstein$ MPC ($k T = 7.8^{+1.4}_{-1.0}$ keV; David et al.\ 1993)
or that determined below from {\it ASCA}
($kT = 8.69^{+0.28}_{-0.30}$ keV; \S~\ref{sec:global_gis}).
An examination of the residuals for this single temperature fit suggested
that the poor fit was due to additional soft X-ray emission.

The poor fit of a single temperature spectrum to the PSPC data and
the previous evidence for a cooling flow at the center of the cluster
suggest that a cooling flow component be added to the model spectrum.
The best-fit single temperature plus cooling flow model is shown in
the second row of Table~\ref{tab:spectra}.
This provided a considerably improved fit (a reduction of
$\Delta \chi^2 \approx 10$ for one additional d.o.f.).
In this model, 53\% of the counts in the PSPC spectrum come from the
cooling flow component.
Because of the soft band of the PSPC, the temperature of the gas
is very poorly defined in this model.
At the 90\% confidence limit, only a lower limit on the temperature
is determined.
As noted before, varying the temperature has little effect on the
soft X-ray spectrum in the single temperature model, and even less
in the cooling flow model (eq.~\ref{eq:cflow}).
The reason for the latter is that most of the softer emission in the
cooling flow model comes from gas at which has cooled to lower temperature,
so that the integral in equation~(\ref{eq:cflow}) is nearly independent of
the upper limit temperature $T$ for $E \ll kT$.

The abundance of heavy elements is also rather poorly determined by
the single temperature plus cooling flow fit to	spectrum, and only
gives a lower limit.
As noted above, the Fe K lines cannot be observed with {\it ROSAT}
to constrain the abundances.
The cooling flow component does produce strong line features within
the PSPC band due to heavy elements, including the L lines of iron
and the K lines of C, N, O, Ne, and Mg.
However, the strengths of these lines are not strongly dependent on
the abundance of the heavy elements, at least if they are all assumed
to vary together and the abundance is not too low.
The reason is that, at temperatures $k T \sim 1$ keV where these
lines are strongest, heavy element line emission dominates the cooling
of the gas.
Thus, if the abundances are increased, the line emission rate is
increased but the cooling time is decreased by nearly the same factor.
In equation~(\ref{eq:cflow}), increasing the abundance increases both
$\Lambda_E ( T' )$ in the numerator and $\Lambda_{tot}$ in the denominator
by approximately the same amount.

On the other hand, the {\it ROSAT} PSPC spectrum does give an excellent
determination of the absorbing column, and a good determination of the
cooling rate $\dot{M}_{cool}$.
As noted above, the absorbing column is reasonably close to the Galactic
value from Stark et al.\ (1992).
The cooling rate also agrees well with previous determinations;
we discuss this in more detail below (\S~\ref{sec:global_both}).

\subsection{{\it ASCA} GIS Global X-ray Spectrum} \label{sec:global_gis}

The {\it ASCA} GIS2 and GIS3 spectra in the photon energy range of
0.5--11 keV were determined for the inner $16'$ of the cluster.

We didn't use the {\it ASCA} SIS0 or SIS1 spectra, because these instruments
did not cover the entire region of the cluster
(Figure~\ref{fig:pspc_image}).
Later, we will find that the cluster has significant spatial variations
in its X-ray spectrum, due to a cooling flow at the cluster center
(\S~\ref{sec:cflow}) and due to a radial temperature gradient
(\S~\ref{sec:spatial_t}).
Thus, it would be difficult to directly compare the SIS spectra with
those from the GIS and PSPC, which cover the same larger region of the cluster.

The results of fits to these spectra are summarized in rows 3 and 4
of Table~\ref{tab:spectra}.
Because of the hard energy band of the GIS spectra, these observations
have strengths and weaknesses which are the opposites of those of the
{\it ROSAT} PSPC.
The GIS provides accurate measures of the ambient gas temperature and
the iron abundance, but do not usefully constrain the soft X-ray absorption
or the cooling rate.

These two spectra were fit first with the single temperature model.
This gave an accurate determination of the temperature and of
the iron abundance, but only an upper limit on the absorbing column.
The fit was adequate, with a $\chi^2$ per d.o.f.\ of 1.08.
Because of the evidence for a cooling flow in A2029, we also tried
a fit which included a cooling flow.
This provided only a marginal improvement in the fit ($\Delta \chi^2 = 2$
for one extra fitting parameter), and accordingly, only an upper limit
on the cooling rate.

\subsection{Combined {\it ASCA--ROSAT} Global Spectral Fits}
\label{sec:global_both}

Because of the complementary strengths and weaknesses of the the
{\it ASCA} and {\it ROSAT} spectra, and the fact that the best-fit
individual spectra gave consistent results (Table~\ref{tab:spectra}),
we determined the best-fit spectrum for the combined data sets
of {\it ROSAT} PSPC and {\it ASCA} GIS2 and GIS3 spectra.
We allowed for a variation in the overall calibration between
{\it ASCA} and {\it ROSAT}, but required that the shape of the
spectrum agree.
The results of fitting single temperature and single temperature plus
cooling flow models to the spectra are shown in rows 5 and 6 of
Table~\ref{tab:spectra}.
The abundances in these fits are essentially the same as those given
by the GIS spectra alone, while the temperatures are slightly lower.
The absorbing columns are similar to those from the PSPC alone, but
with smaller uncertainties.
The cooling rate in the CF model is basically that same as that given
by the PSPC alone.

\begin{figure*}[tbh]
\vskip4truein
\includegraphics{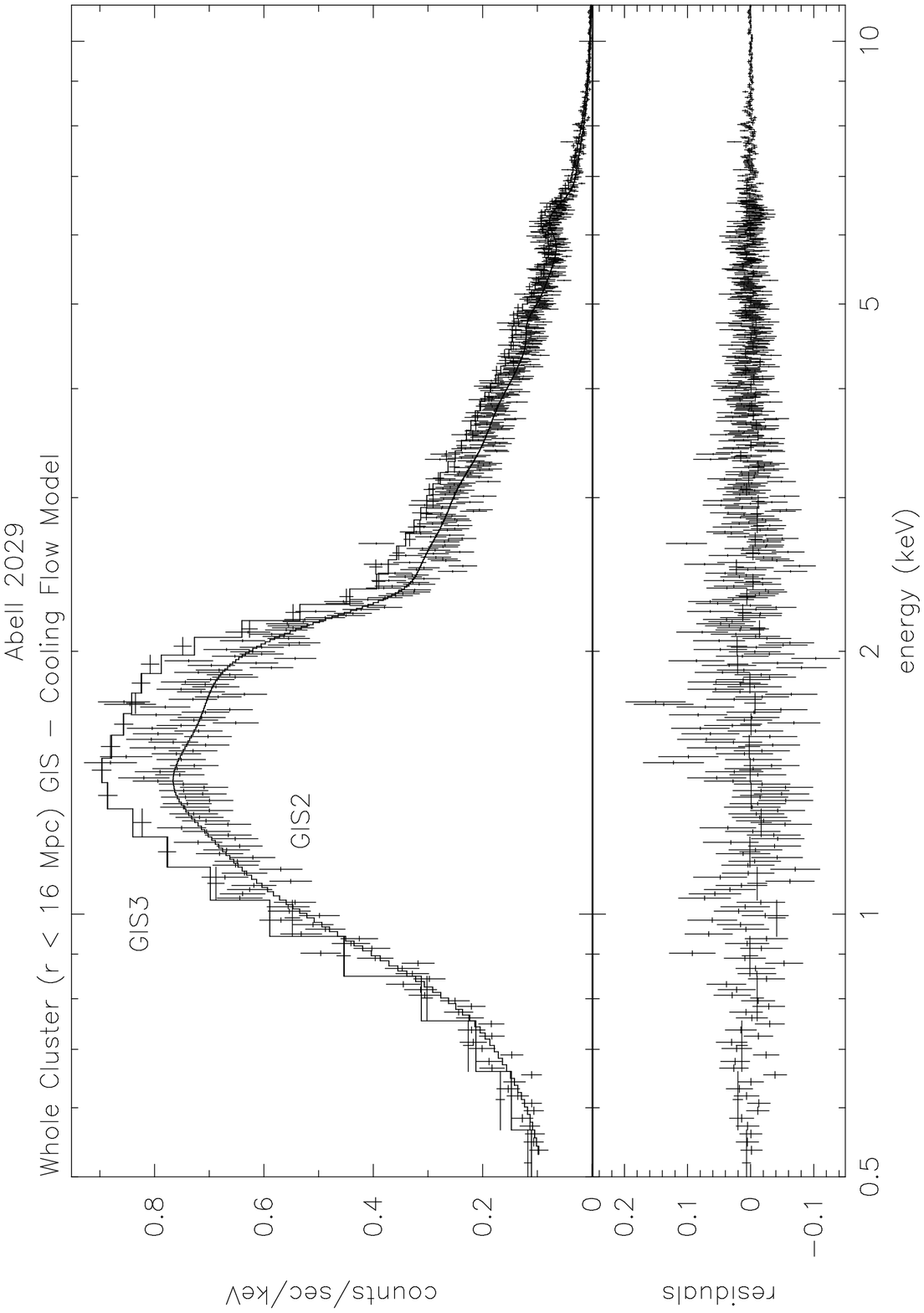}
\caption
{The global {\it ASCA} GIS2 and GIS3 X-ray spectra for A2029
($r \le 16'$) as a function of the measured photon
energy are shown is the upper panel.
The crosses give the data points with 1-$\sigma$ error bars,
while the histogram is the best single temperature plus cooling flow
model fit to the PSPC, GIS2, and GIS3 spectra
(row 6 in Table~\protect\ref{tab:spectra}).
The width of the data points or histogram steps is the width of
the energy channels used to accumulate the data.
The upper spectrum is for the GIS3, while the lower spectrum
is for GIS2.
The GIS3 has larger energy bins because the bit error in memory
during the observation reduced the energy resolution
(\S~\protect\ref{sec:obs_asca}).
The lower panel gives the residuals to the fit (in counts/sec/keV).
}
\label{fig:whole_asca_spect}
\end{figure*}

\begin{figure*}[tbh]
\vskip4truein
\includegraphics{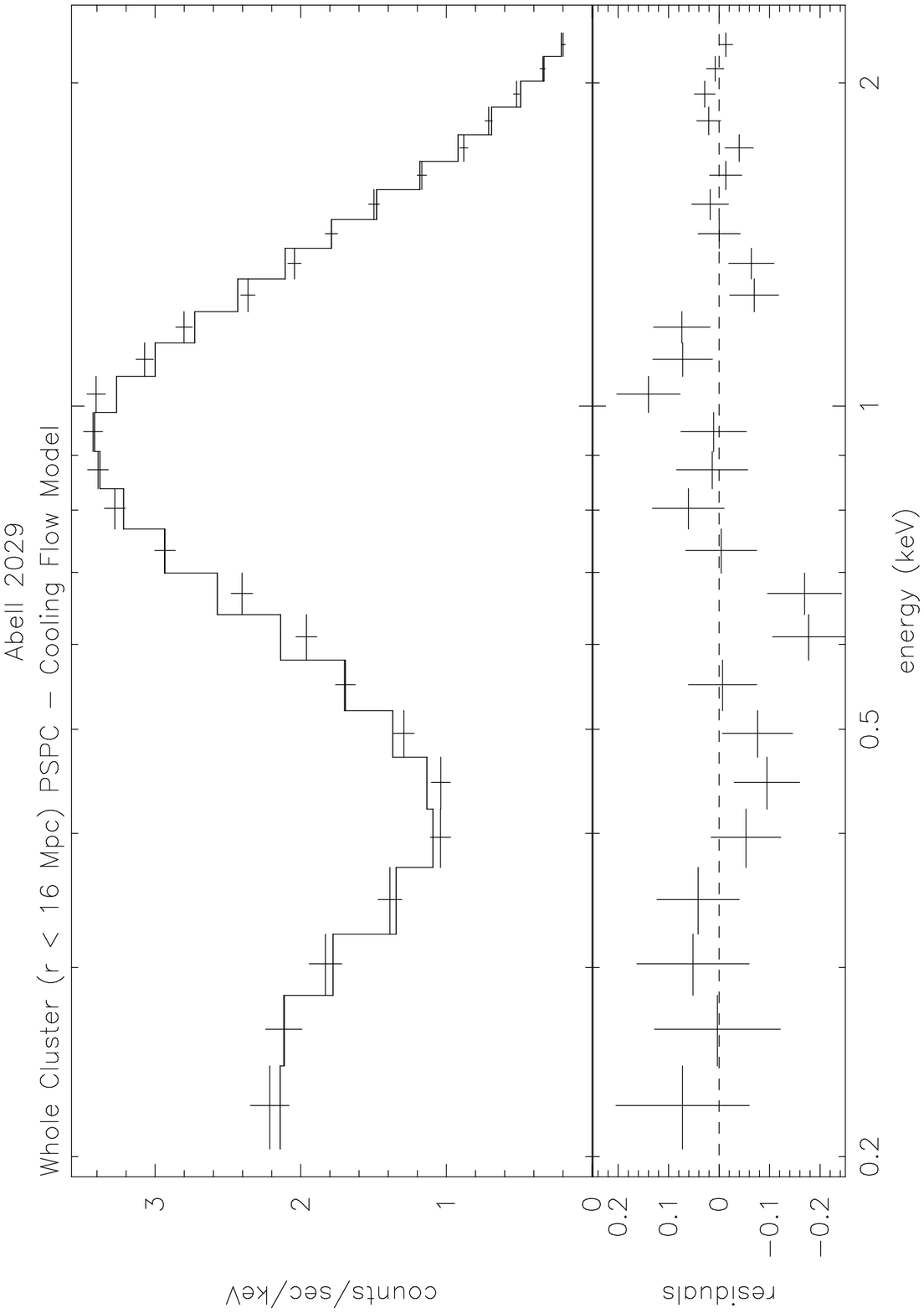}
\caption{The global {\it ROSAT} PSPC X-ray spectrum for A2029
($r \le 16'$) as a function of the measured photon
energy is shown is the upper panel.
The crosses give the data points with 1-$\sigma$ error bars,
while the histogram is the best single temperature plus cooling flow
model fit to the PSPC, GIS2, and GIS3 spectra.
(row 6 in Table~\protect\ref{tab:spectra}).
The width of the data points or histogram steps is the width of
the energy channels used to accumulate the data.
The lower panel gives the residuals to the fit (in counts/sec/keV).
}
\label{fig:whole_pspc_spect}
\end{figure*}

The addition of a cooling flow improves the fit significantly
($\Delta \chi^2 = 38$ for one extra fitting parameter).
This is due almost entirely to the effect of this component on the
fit to the PSPC spectrum.
Because the single temperature plus cooling flow model provides
fairly good fit to both the {\it ROSAT} and {\it ASCA} data, we
will adopt the spectrum in this fit as our fiducial global cluster
spectrum for A2029.
The upper panel in
Figure~\ref{fig:whole_asca_spect} shows the observed global {\it ASCA}
GIS2 and GIS3 spectra and this best-fit model;
the lower panel shows the residuals to the fit.
Figure~\ref{fig:whole_pspc_spect} shows the {\it ROSAT} PSPC spectrum,
model, and residuals.

The hydrogen absorbing column required by this fit is in reasonable
agreement with the Galactic value of $N_H = 3.07 \times 10^{20}$ cm$^{-2}$
from Stark et al.\ (1992).
The temperature in the single temperature plus cooling flow model is
higher than the value of $k T = 7.8^{+1.4}_{-1.0}$ keV found by
David et al.\ (1993) with the {\it Einstein} MPC, although the error bars
overlap.
However, David et al.\ fit a single temperature model to their spectrum.
If one compares to the single temperature joint fit to the {\it ASCA}
and {\it ROSAT} data (row 5 in Table~\ref{tab:spectra}), the agreement
is quite good.
The total cooling rate is in remarkably good (and probably fortuitous)
agreement with the values of
$\dot{M}_{cool} = 366 h_{50}^{-2} \, M_\odot$ yr$^{-1}$
derived with the {\it Einstein} IPC (Arnaud 1989), and
$\dot{M}_{cool} = 370 h_{50}^{-2} \, M_\odot$ yr$^{-1}$
derived from the {\it ROSAT} HRI (Sarazin et al.\ 1992).
Both of these determinations were based on the surface brightness profile
of the cluster, rather than the X-ray spectrum.
The cooling rate was also derived using the X-ray spectrum from the
{\it Einstein} SSS by White et al.\ (1991).
Their value of
$\dot{M}_{cool} = 513^{+304}_{-247} h_{50}^{-2} \, M_\odot$ yr$^{-1}$
is consistent with ours within their very large uncertainties.

The X-ray flux in these models is
$F_X$ (2--10 keV) $= (1.00 \pm 0.01) \times 10^{-10}$ ergs cm$^{-2}$ s$^{-1}$.
The uncertainty shown is statistical, and is dominated by the
good statistics in the {\it ASCA} GIS spectra.
However, the systematic uncertainty in the absolute calibration of
these {\it ASCA} spectra is much larger.
This flux is considerably larger than that found by
David et al.\ (1993),
$F_X$ (2--10 keV) $= 0.75 \times 10^{-10}$ ergs cm$^{-2}$ s$^{-1}$,
using the {\it Einstein} MPC, which had a larger field of view than
{\it ASCA}.

\section{SPATIAL VARIATION OF THE X-RAY SPECTRUM} \label{sec:spatial}

The {\it ASCA} observations were used to study the spatial variation in
the X-ray spectrum, and derive the radial variation in the gas
temperature
and in the abundances.
Both SIS and GIS data were used.
The determination of the spatial variation in the spectrum is
complicated by the broad, energy-dependent Point-Spread-Function (PSF)
of the {\it ASCA} mirrors.
We used the technique of Markevitch et al.\ (1996) to correct for the
redistribution of X-ray photons produced by the {\it ASCA} PSF.
A set of ``model'' regions on the sky were selected, and we assumed that
the shape of the X-ray spectrum was constant within each of these regions.
A corresponding set of ``image'' regions were defined in the
focal plane of {\it ASCA}.
Generally, the image regions and model regions were identical, except
that the outermost model region was larger than the outermost image region.
This was done to avoid the unphysical effects of assuming that the cluster
emission ended abruptly at the outer edge of the last image region.
Also, the image regions were restricted to the area within $18^\prime$
of the center of each of the GIS detectors.

The observed {\it ASCA} X-ray spectra were accumulated within each
image region.
For a given set of spectral parameters for the model regions, we determined
the predicted spectra in the image regions, including the detector response
and the redistribution of photons among image regions by the energy-dependent
{\it ASCA} PSF.
The parameter of the spectra in the model regions were varied until
the best-fit model (minimum $\chi^2$) for the observed spectra in the image
regions was found.
An annealing technique (Press et al.\ 1992) was used to avoid false minima.
We were careful to use spectral bins which were large enough that the
Gaussian approximation to Poisson uncertainties was valid, so that the
$\chi^2$ test could be applied.

The energy-dependent {\it ASCA} PSF was approximated by interpolation
between GIS images of Cyg X-1 taken at various locations in the
focal plane
(Takahashi 1995).
The PSF is poorly known at photon energies below 2 keV.
For photon energies between 2.0 and 2.5 keV, the effective area
of {\it ASCA} is affected by a poorly calibrated absorption edge from gold
in the mirrors.
Thus, we have chosen to fit only photon energies above 2.5 keV
in the spatial analysis.
The lower energy cutoff should not affect the determinations of
the high temperatures found for the gas.
Excluding lower energies does reduce the sensitivity of {\it ASCA} to
the central cooling flow
(\S~\ref{sec:cflow}),
but also reduces the calibration problems encountered with
{\it ASCA} at low energies (mainly below 1 keV; \S~~\ref{sec:cflow_general}).
A relative systematic PSF uncertainty (5\% for circularly symmetric image
regions, 15\% for other region shapes) was added to the statistical
uncertainties.

The instrumental contributions to the PSF are different for the SIS and
GIS.
For modeling of the SIS data, the GIS Cyg X-1 calibration images were
corrected for the energy dependence of the intrinsic GIS detector blurring
by additional smoothing resulting in a final constant resolution (Gaussian
$\sigma=0\farcm5$). The cluster SIS data also were smoothed to the same
resolution.

Because of its superior spatial resolution, the {\it ROSAT} PSPC image
(Fig.~\ref{fig:pspc_image}) was used to determine the distribution
of projected emission measure within the cluster.
Errors due to photon statistics and the background correction in the
{\it ROSAT} PSPC image where included in the spectral analysis.
At each step in the iteration to find the best-fit {\it ASCA} spectra,
the spectrum in each model region was used to determine the emission
within the band of the {\it ROSAT} PSPC image including the
{\it ROSAT} PSPC response, and the surface brightness in the PSPC
image was converted into a distribution of emission measure on the
sky.
Because there is some uncertainty in the overall {\it ROSAT} to
{\it ASCA} calibration, we allow this to vary, and only use the
{\it ROSAT} PSPC image to fix the relative emission measure variation
across the sky.
This emission measure distribution and the spectral parameters were
used to determine the predicted {\it ASCA} spectra within the image
regions, including the redistribution by the PSF.
Because a small miss-alignment of the {\it ROSAT} and {\it ASCA} images
can affect the spectral fit,
it is important to align the {\it ROSAT} and {\it ASCA} data accurately.
We did this, but we also vary the offset between the two satellite data
sets, and include an estimate of the resulting uncertainties in the alignment
on our spectral fits.

The use of the {\it ROSAT} PSPC image to constrain the emission measure
distribution in the {\it ASCA} only works well if the same spectral
component provides the bulk of the emission in these two disjoint
spectral bands
(0.5--2.0 keV for the {\it ROSAT} image, and
2.5--11 keV for the {\it ASCA} data).
This is probably a reasonable assumption in the outer parts of A2029,
but not in the inner cooling flow region.
There, most of the {\it ROSAT} photons come from cooling gas,
while much of the {\it ASCA} spectrum is contributed by hotter,
ambient gas.
To avoid errors due to this, we free the normalization of the
central {\it ASCA} spectrum relative to the outer regions
during the spectral fitting.

Two bright X-ray point sources are seen on the {\it ROSAT} PSPC image
(Fig.~\ref{fig:pspc_image}) at about 4 and 9 arcmin from the cluster
center.
A region around each of these sources was excluded from the 
{\it ASCA} cluster spectra.
However, because of the broad {\it ASCA} PSF, some emission from each of
these point sources may be included in the adjoining cluster image
regions.
Thus, we accumulate the {\it ASCA} X-ray spectrum from the image regions
around each of these sources, and simultaneously fit a power-law spectrum
to each of these sources.
(Thermal spectra were found to provide poorer fits than power-law
spectra.)
We found that contamination from these sources only has a small effect on
the derived spectral properties of the cluster regions except for the
outermost cluster regions.
Because these sources may have varied between the time of the {\it ROSAT}
and {\it ASCA} observations, we allow the overall normalization of the flux
of these sources to vary.
In fact, the {\it ASCA} and {\it ROSAT} fluxes of source A
(Fig.~\ref{fig:pspc_image}) agree within the uncertainties when corrected for
the different spectral bands, but source B is about 3 times fainter in
the {\it ASCA} observation than in the {\it ROSAT} data.
The photon power-law indices of sources A and B were
$-1.49^{+0.33}_{-0.42}$ and $-1.11^{+0.37}_{-0.34}$,
respectively, in the best-fit models.

The uncertainties in derived spectral quantities are 90\% confidence regions
from 200 Monte Carlo simulations of fitting the data with random
statistical and systematic uncertainties added.

\subsection{Radial Temperature Gradient} \label{sec:spatial_t}

The radial temperature variation in A2029 was determined by accumulating
the {\it ASCA} SIS and GIS spectra in four adjoining annuli with
boundary radii of 0, 1.5, 5, 10, and 16 arcmin.
The outer radius of the outermost model annulus was taken to be 20 arcmin.
The spectra were grouped into eight spectral channels with boundary
photon energies of
2.5, 3.0, 3.5, 4.0, 5.0, 6.0, 7.0, 8.5, 11.0
keV in the GIS detectors.
In the SIS detectors, the last two of these channels were combined
into a single channel from 7 to 11 keV.
For the purposes of determining the temperature variation, the
abundances of the heavy elements were fixed at
0.40 of the solar value, which was the best-fit abundance for the
entire cluster spectrum (\S~\ref{sec:global}).

We initially tried to fit the spectra assuming a single temperature
model for the spectrum in all of the cluster annuli.
However, this did not provide a very good fit to the central
circular region.
Even when we freed the normalization of this region (i.e.,
no longer set the normalization of this region relative to the
others from the {\it ROSAT} PSPC image), the fit was only improved
very slightly.
Examination of the residuals indicated that the model was unable to
provide enough soft flux in the central region.
Previous X-ray spectral
(e.g., White et al.\ 1991)
and
spatial observations
(e.g., Sarazin et al.\ 1992),
and the central {\it ROSAT} PSPC and {\it ASCA} spectra presented in this
paper (\S~\ref{sec:cflow}),
all indicate that A2029 contains a cooling flow in its central regions.
Therefore, we added a cooling flow component to the central circle
region in all of our spatially-resolved spectral fits.
The gas was assumed to cool from the ambient temperature in this
region (the temperature of the Raymond-Smith component), and have
the same abundances as the intracluster gas.
The addition of this cooling flow component lowered $\chi^2$ by
32 for only one additional parameter.
This model did provide an acceptable fit.

The best-fit spectral parameters for this model for the cluster annuli
are shown in rows 7--10 of Table~\ref{tab:spectra}.
Note that the best-fit value of $\chi^2$/d.o.f.\ is significantly
less than unity.
This is due to the inclusion of systematic uncertainties in the evaluation
of $\chi^2$, as discussed above.
As a result of this, we do not give the probabilities associated with
these fits.

The best-fit temperatures in annuli are also plotted in
Figure~\ref{fig:asca_temp}.
There is some evidence for a decrease in the temperatures with radius.
This trend was found in all of the fits to the {\it ASCA} spectra, including
the fits which did not correct for the two unrelated sources
or include the effects of the central cooling flow.
To assess the significance of the temperature gradient, the spectra in
annuli were fit assuming that the temperature was the same at all radii.
The horizontal dashed line in Figure~\ref{fig:asca_temp} gives this best-fit
isothermal model for all of the spatially-resolved spectra ($kT = 8.6$ keV).
Note that this temperature is slightly different than the best-fit
temperature for the whole cluster (also including the effects of a cooling
flow), $kT = 9.35$ keV, for several reasons.
First,
This was a much worse fit to the spectra.
Based on the F-test and the simulations of the uncertainties in the fits,
this isothermal fit can be rejected at the $>$ 96\% confidence level.
To parameterize the temperature gradient, we also fit the spectra in
annuli assuming that the temperature was a linear function of the
radius.
The slanting dashed line in Figure~\ref{fig:asca_temp}
is the best-fit linear function of radius,
\begin{equation} \label{eq:t_linear}
\left( \frac{kT}{1 \, {\rm keV}} \right)
= 9.49 - 0.322 \left( \frac{r}{1 \, {\rm arcmin}} \right) \, .
\end{equation}
It is an improvement over the isothermal temperature distribution, with
a reduction in $\chi^2$ of 9 for only one additional parameter.

\centerline{\null}
\vskip2.5truein
\includegraphics{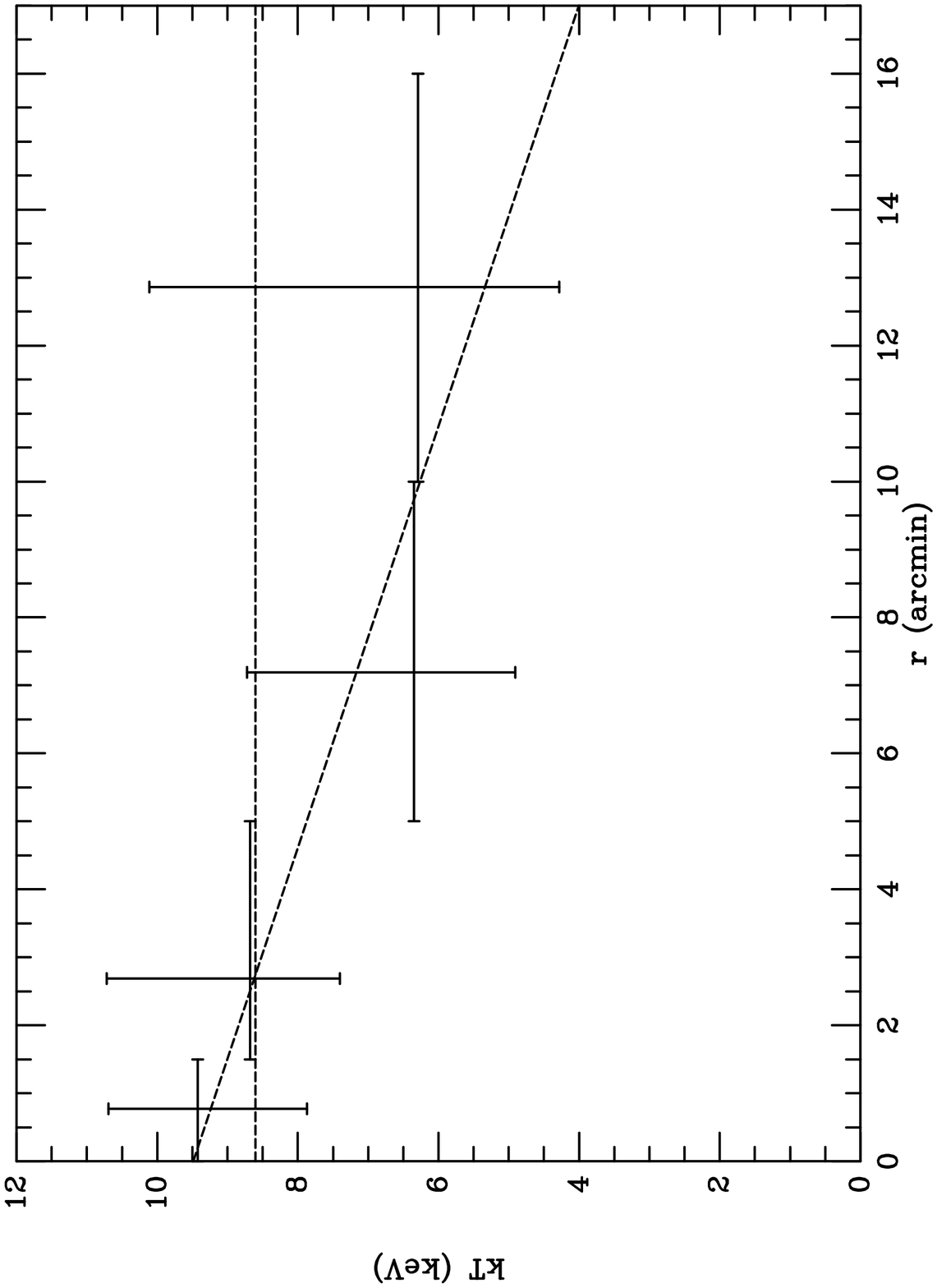}
\figcaption
{The temperature of the intracluster gas in A2029 as a
function of radius as derived from the {\it ASCA} spectra.
The vertical error bars give the 90\% confidence region for the
temperature.
The horizontal error bars show the radii of the annuli over which the
spectra were accumulated.
The temperature is plotted at the median radius of the X-ray emission
in that annulus.
The horizontal dashed line gives the best-fit single temperature
fit to the individual annular spectra.
The sloping dashed line is the best-fit linear function of the
radius (eq.~\protect\ref{eq:t_linear}).
The spectral model used for the central bin includes a cooling flow,
and the temperature plotted is the upper limit temperature, which we
assume is that of the ambient gas in this region.
\label{fig:asca_temp}}

\vskip0.2truein

Fits to the GIS data alone showed the same declining trend of temperatures
with radius and consistent values for the temperatures.
In fits with the SIS data alone, the temperatures in the inner two
annuli were similar to those in the combined fit.
The temperatures in the outer two annuli were very poorly determined,
due to the small portion of these regions included on the SIS chips
(Figure~\ref{fig:pspc_image}).
However, the temperatures were consistent within the uncertainties, which were
very large in the outer two annuli.

\subsection{Radial Abundance Variation} \label{sec:spatial_ab}

We also derived the abundances in the same radial rings which were
used in the previous section.
The values of the derived abundances are given in Figure~\ref{fig:asca_abund}.
Unfortunately, the uncertainties in the individual abundances are quite large,
and only upper limits are found for the outer two annuli.
The is not really due to the statistical uncertainties on individual spectra.
Instead, it is largely a consequence of the energy dependent PSF of
{\it ASCA}.
The PSF is particularly wide at high energies, so the Fe K line complex,
which is the primary spectral feature determining the abundance, can be
strongly affected.
When combined with the very strong central surface brightness of the
Fe K line complex in this cooling flow cluster, this implies that
there is a very large correction for scattering of the central emission
in the Fe K line spectra of the outer regions.
This results in very large uncertainties for the outer abundances, and causes
the uncertainties in the outer abundances to be anticorrelated with the
uncertainties in the central abundance.

\centerline{\null}
\vskip2.5truein
\includegraphics{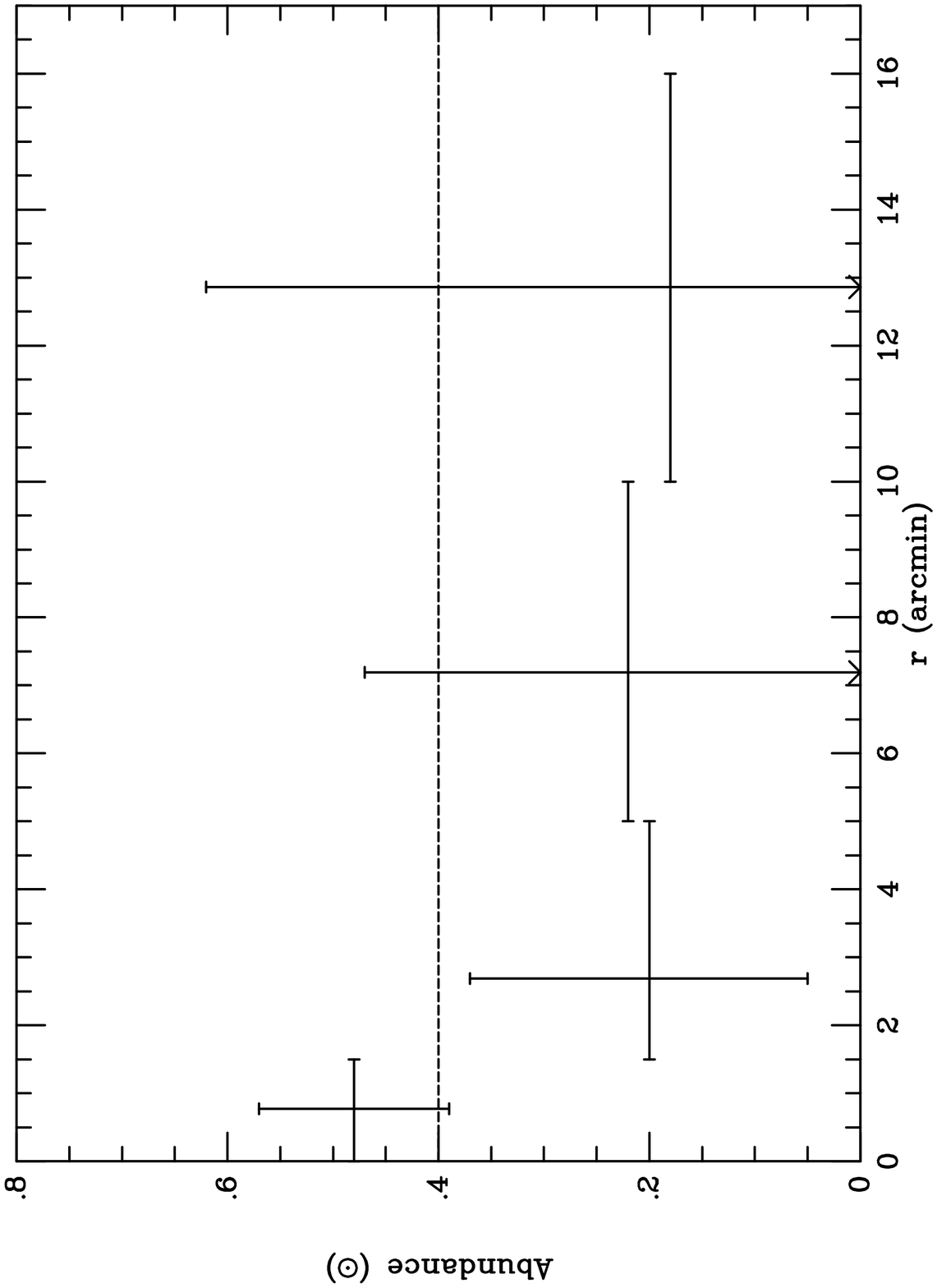}
\figcaption
{The iron abundance of the intracluster gas in A2029 as a
function of radius as derived from the {\it ASCA} spectra.
The vertical error bars give the 90\% confidence region for the
abundances.
The horizontal error bars show the radii of the annulus over which the
spectra were accumulated.
The abundance is plotted at the median radius of the X-ray emission
in that annulus.
The dashed horizontal line is the best-fit single abundance value for
the cluster.
There is no statistically significant variation of the abundance with
radius.
\label{fig:asca_abund}}

\vskip0.2truein

Figure~\ref{fig:asca_abund} shows some slight evidence for either a
decline in the iron abundance with radius or a higher abundance in the
center cooling flow region than in the rest of the cluster.
However, the difference between the spectral fit with individual abundances
and a fixed overall abundance is not statistically significant
(a decrease in $\chi^2$ of 3.3 for three additional parameters).
One finds that the hypothesis that the abundance is constant can only
be rejected at the 51\% confidence level (i.e., not at all).
For this reason, we have assumed that the abundance is constant at the
value set by the global cluster spectrum in the determinations of the
cluster temperature variation.

\subsection{Azimuthal Temperature Variation} \label{sec:spatial_slice}

We also determined the azimuthal variation in the X-ray temperature
in A2029.
As shown in Figure~\ref{fig:pspc_image}, the X-ray emission in A2029
is elongated at a position angle ($PA$) of approximately $25^\circ$
(measured from the north to the east).
We divided the region of the cluster between radii of 3$'$ and 16$'$
into four annular sections, each $90^\circ$ wide, with the first centered
on $PA = 25^\circ$.
The temperature was determined independently for the central 3 arcmin
radius circle and for each of these annular sections.
These regions are shown in Figure~\ref{fig:pspc_slices}.
As before, regions were used to isolate the two point sources
A and B.
The abundance was fixed at 0.40 of solar, the best-fit value for
the cluster spectrum derived above.
The relative systematic PSF uncertainty was increased to
15\% because of the shape of the regions used.

\centerline{\null}
\vskip2.7truein
\includegraphics{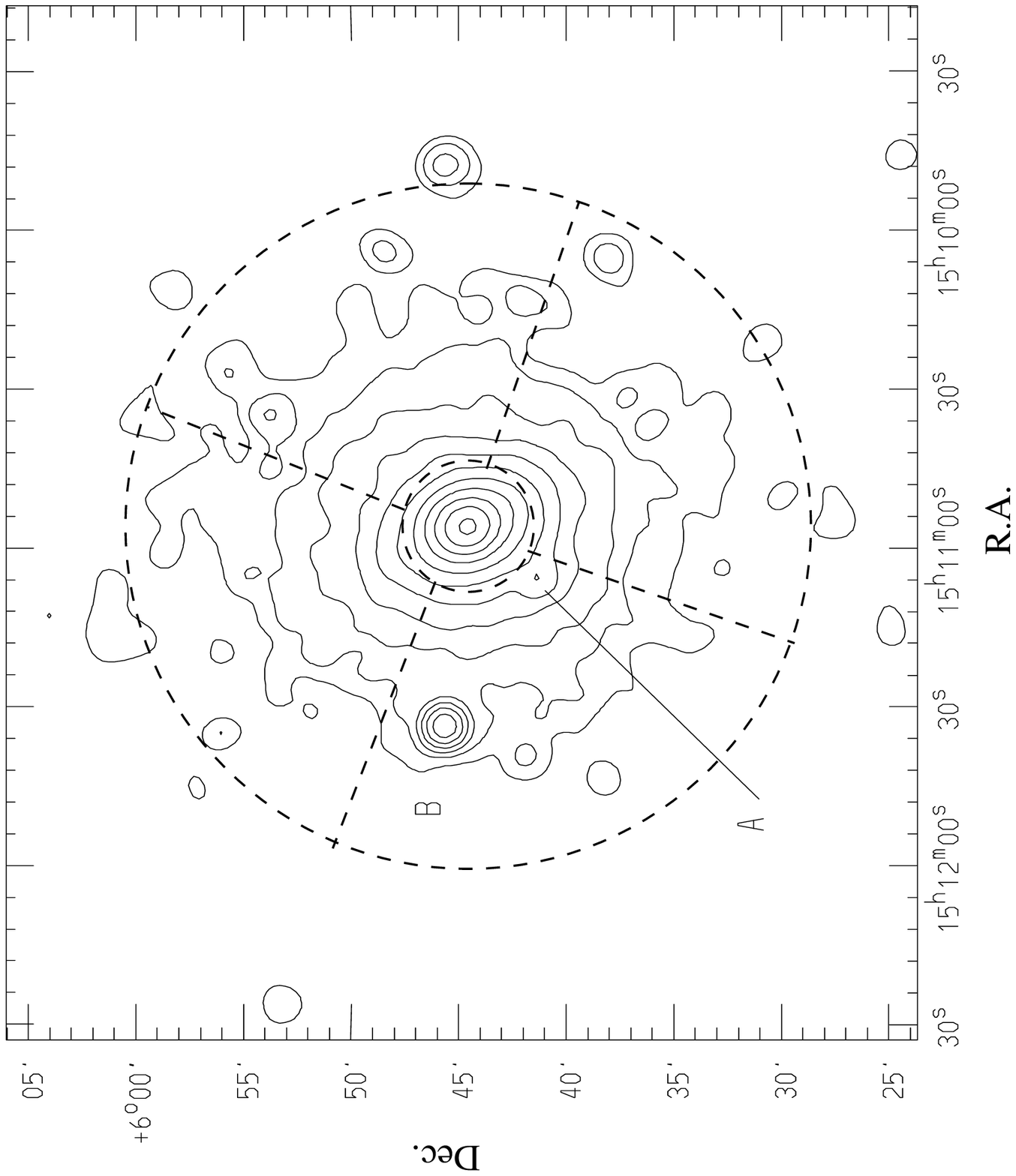}
\figcaption
{The annular sections used to derive the azimuthal
variation of the temperature in A2029.
The contours give the {\it ROSAT} PSPC surface brightness as
in Figure~\protect~\ref{fig:pspc_image}.
The outer dashed circle has a radius of 16 arcmin.
Only regions interior to this circle were used to derive the
global or spatially resolved X-ray spectra.
The inner dashed circle has a radius of 3 arcmin;
the spectrum interior to this regions was used to determine
the properties of the cooling flow in the cluster
(\S~\protect\ref{sec:cflow}).
The radial dashed lines delimit the 4 regions used to determine
the azimuthal temperature variation
(\S~\protect\ref{sec:spatial_ab}).
\label{fig:pspc_slices}}

\vskip0.2truein

The resulting temperatures for the four regions are listed in
rows 11-14 of Table~\ref{tab:spectra}, and are shown as a function of
the position angle in Figure~\ref{fig:tslices}.
There is some evidence that the gas is hotter to the northwest,
and cooler to the southeast.
We also fit a single temperature to all four sectors;
this is shown as a dashed horizontal line in Figure~\ref{fig:tslices}.
Formally, the measured temperatures disagree with this single temperature
fit at almost the 90\% confidence level.
However, the presence of the two point sources within the cooler sector
and increased systematic uncertainties in fitting the spectra of regions
which are not azimuthally symmetric argues for caution in interpreting
these results.
We conclude that the uniformity of the gas temperature with azimuthal
angle cannot be ruled out at the greater than 90\% confidence level.

\section{CENTRAL COOLING FLOW SPECTRA} \label{sec:cflow}

\subsection{General Properties} \label{sec:cflow_general}

We also extracted the spectrum of the inner region of the cluster to
study the properties of the cooling flow in more detail.
The SIS and GIS spectra were extracted for the inner 3 arcmin
(360 kpc) in radius of the cluster.
This region just fits on the S0C1 and S1C3 chips, which are the
``best'' chips on the two instruments and the ones normally used to
determine the spectra of point sources.
We note that our treatment of the central cooling flow spectra differed
from that of the previous analysis of the global cluster spectrum
(\S~\ref{sec:global}) or of the spatially resolved spectra
(\S~\ref{sec:spatial}) in several ways.
First, we included the spectra from the two SIS instruments;
these were excluded in the analysis of the global spectrum because
the cluster extended beyond the field of view of the SIS.
Second, we include the lower energy channels in the spectra,
which were not included in the spatial analysis in \S~\ref{sec:spatial}.
Because our interest here is in the soft X-ray emitting cooling flow,
we extend the spectral analysis down to energies of 0.35 keV
in the SIS and 0.55 keV in the GIS.

\centerline{\null}
\vskip2.5truein
\includegraphics{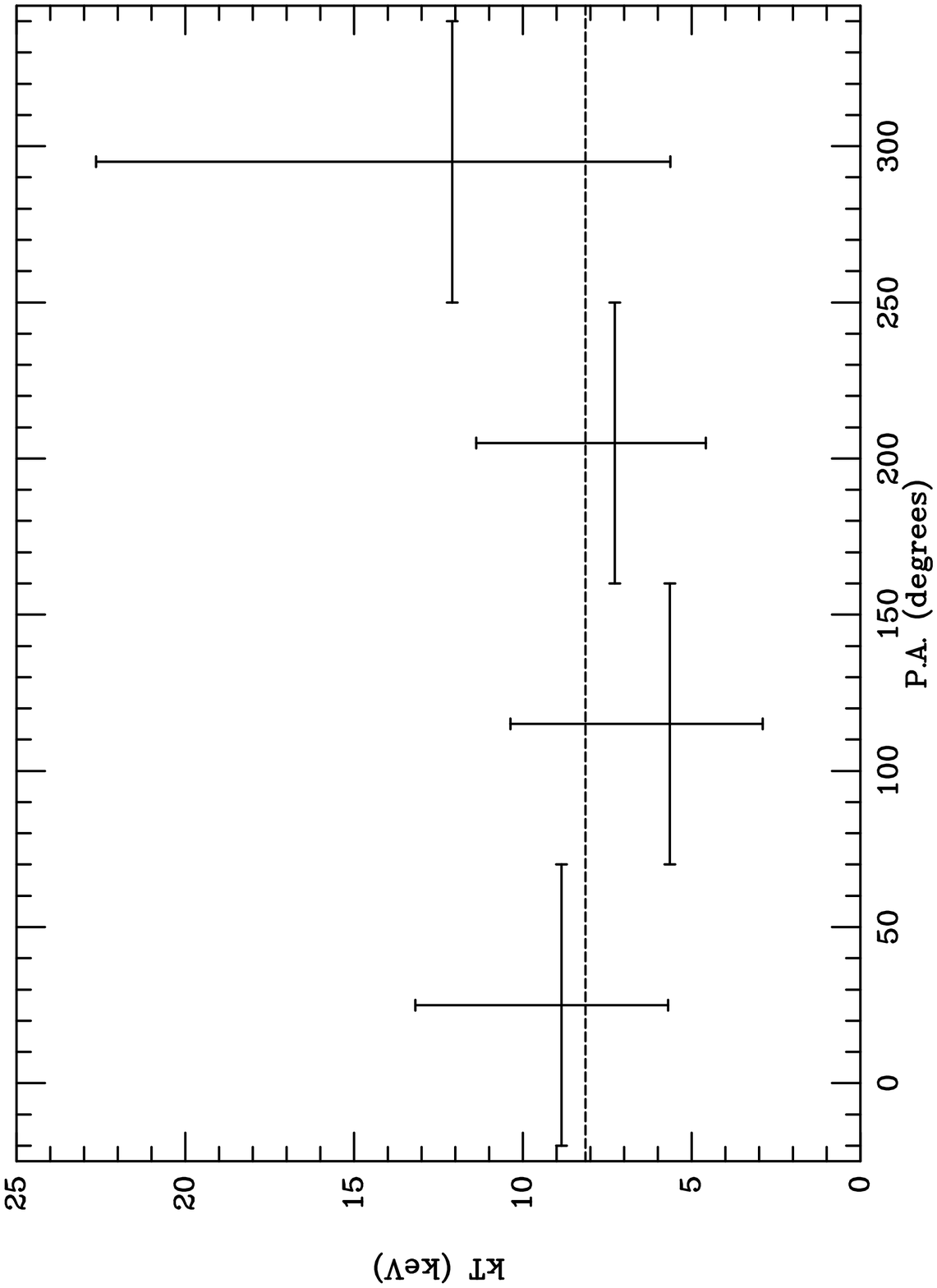}
\figcaption
{The temperature of the gas in A2029 at radii between
3 and 16 arcmin as a function of the position angle around the center of the
cluster.
The position angle $PA$ is measured from north to the east.
The dashed horizontal line shows the best-fit single temperature model
for the same spectra.
The error bars are the 90\% confidence regions.
\label{fig:tslices}}

\vskip0.2truein

The energy-dependent PSF of {\it ASCA} can cause problems in determining
the spectrum of any component of an extended X-ray source like a cluster.
In \S~\ref{sec:spatial}, we corrected for these effects while simultaneously
fitting simple models to the X-ray spectrum in several regions of the
the cluster.
Unfortunately, we will fit more complicated models to the spectrum of the
cooling flow, and such a simultaneous fit is not practical with these
models.
Also, to study the cooling flow, it is important to extend the spectral
analysis down to lower energies where the PSF is not well-understood.
We expect that the problems associated with the energy-dependent PSF will
be much smaller for the cooling flow region than for the outer cluster
regions, because of the large surface brightness peak associated
with the cooling flow.
To test this, we have determined the contribution of outer X-ray
emission scattered into the 3 arcmin circle from outer regions,
assuming the {\it ROSAT} PSPC image for the cluster and the best-fit
spectral parameters as a function of position from \S~\ref{sec:spatial}.
In the GIS, we find that scattered light from outer regions contributes
roughly 8\% to the flux in the central region.
The values are somewhat smaller for the SIS.
There is also an energy-dependent loss of flux from the central region
due to the PSF.
During the spectral analysis, the program {\sc ascaarf} corrects for this
effect.

We also extracted the {\it ROSAT} PSPC spectrum from the same region.
All of the spectra were binned to insure that there were at least 20 counts
per spectral channel, so that the $\chi^2$ test should give reliable results.

Single temperature model fits to the GIS, SIS, and PSPC spectra are shown
in rows 15--17 of Table~\ref{tab:spectra}.
Individually, these are probably acceptable fits.
However, we could not find an acceptable common fit to all of the spectra.
There are two inconsistencies between the fits to the different
instruments.
First, the absorbing column required by the SIS spectrum was about twice
that found with the PSPC or the GIS.
Second, the temperatures found by each of the instruments disagreed
at a level beyond the uncertainties.
It is interesting that the temperature increased with the hardness of the
spectral response of the instrument.
This suggests that multiple temperature components are present, and that
each instrument shows the gas which emits more strongly in its spectral
band.

The three instruments did give consistent values for the heavy element
abundance, and the values for the central spectra are in good agreement
with those determined for the entire cluster.
This agrees with our previous conclusion in \S~\ref{sec:spatial_ab}
that the abundances are nearly constant within A2029.

\begin{figure*}[tbh]
\vskip4truein
\includegraphics{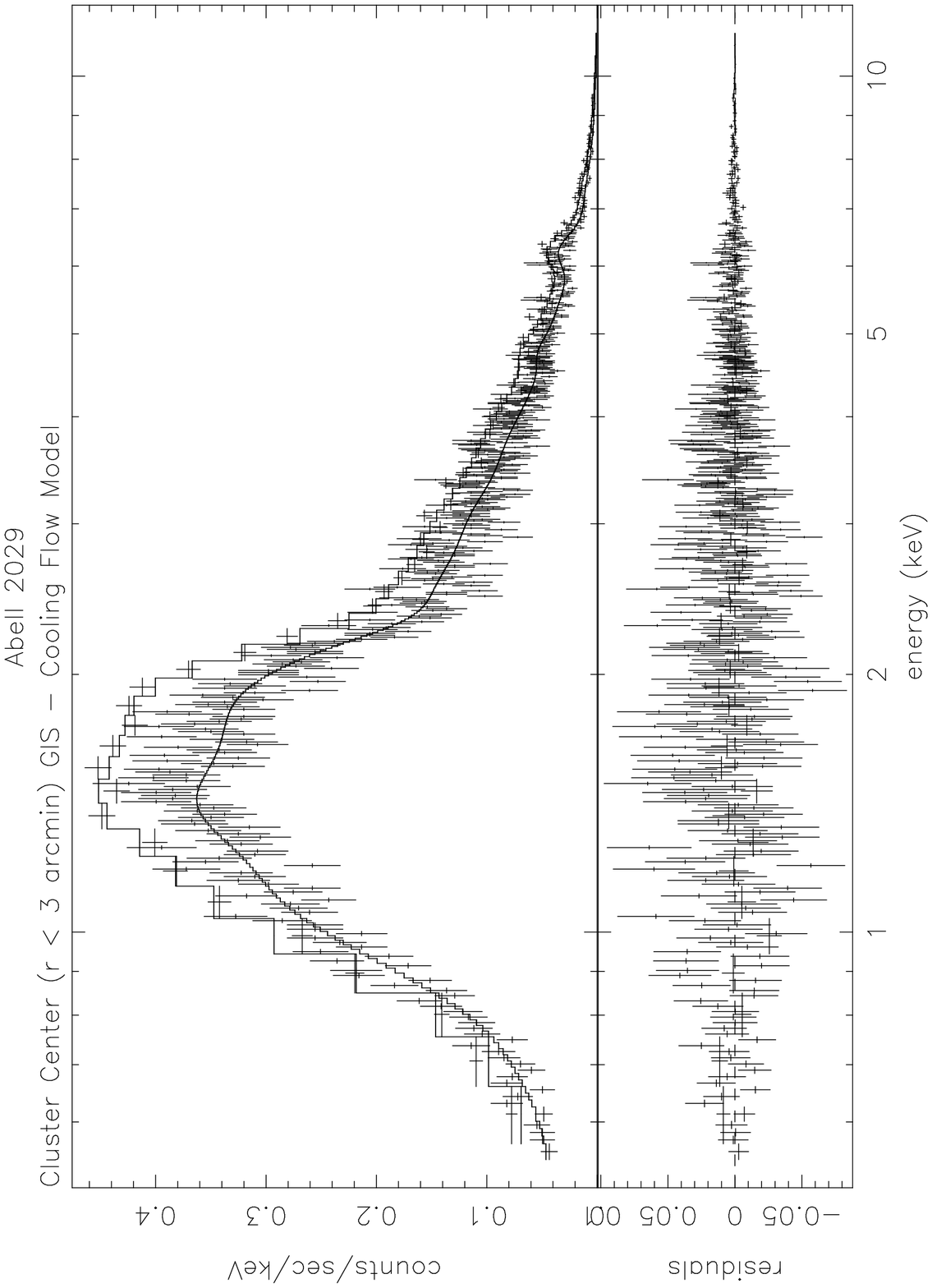}
\caption{The {\it ASCA} GIS2 and GIS3 X-ray spectra for the
central cooling flow region for A2029 ($r \le 3'$) as a function of the
measured photon energy are shown in the upper panel.
The crosses give the data points with 1-$\sigma$ error bars,
while the histogram is the best single temperature plus cooling flow
model for the GIS spectra
(row 18 in Table~\protect\ref{tab:spectra}).
The width of the data points or histogram steps is the width of
the energy channels used to accumulate the data.
The upper spectrum is for the GIS3, while the lower spectrum
is for the GIS2.
The GIS3 has larger energy bins because the bit error in memory
during the observation reduced the energy resolution
(\S~\protect\ref{sec:obs_asca}).
The lower panel gives the residuals to the fit (in counts/sec/keV).
}
\label{fig:center_gis_spect}
\end{figure*}

\begin{figure*}[tbh]
\vskip4.1truein
\includegraphics{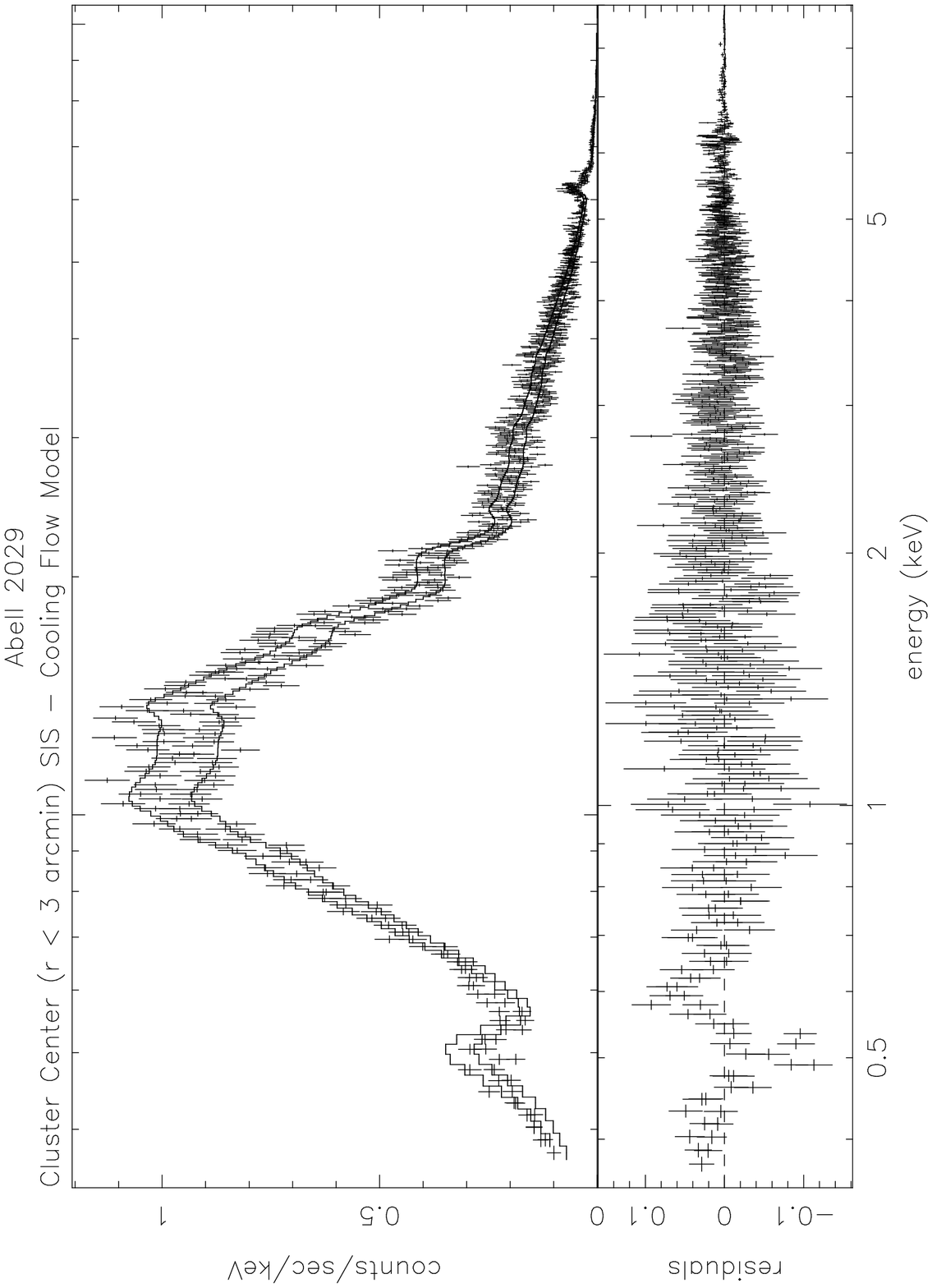}
\caption{The {\it ASCA} SIS0 and SIS1 X-ray spectra for the
central cooling flow region for A2029 ($r \le 3'$).
The notation is the same as for Figure~\protect\ref{fig:center_gis_spect}.
The histogram is the best single temperature plus cooling flow
model for the SIS spectra
(row 19 in Table~\protect\ref{tab:spectra}).
The upper spectrum is for the SIS0, while the lower spectrum
is for SIS1.
}
\label{fig:center_sis_spect}
\end{figure*}

\begin{figure*}[tbh]
\vskip4.1truein
\includegraphics{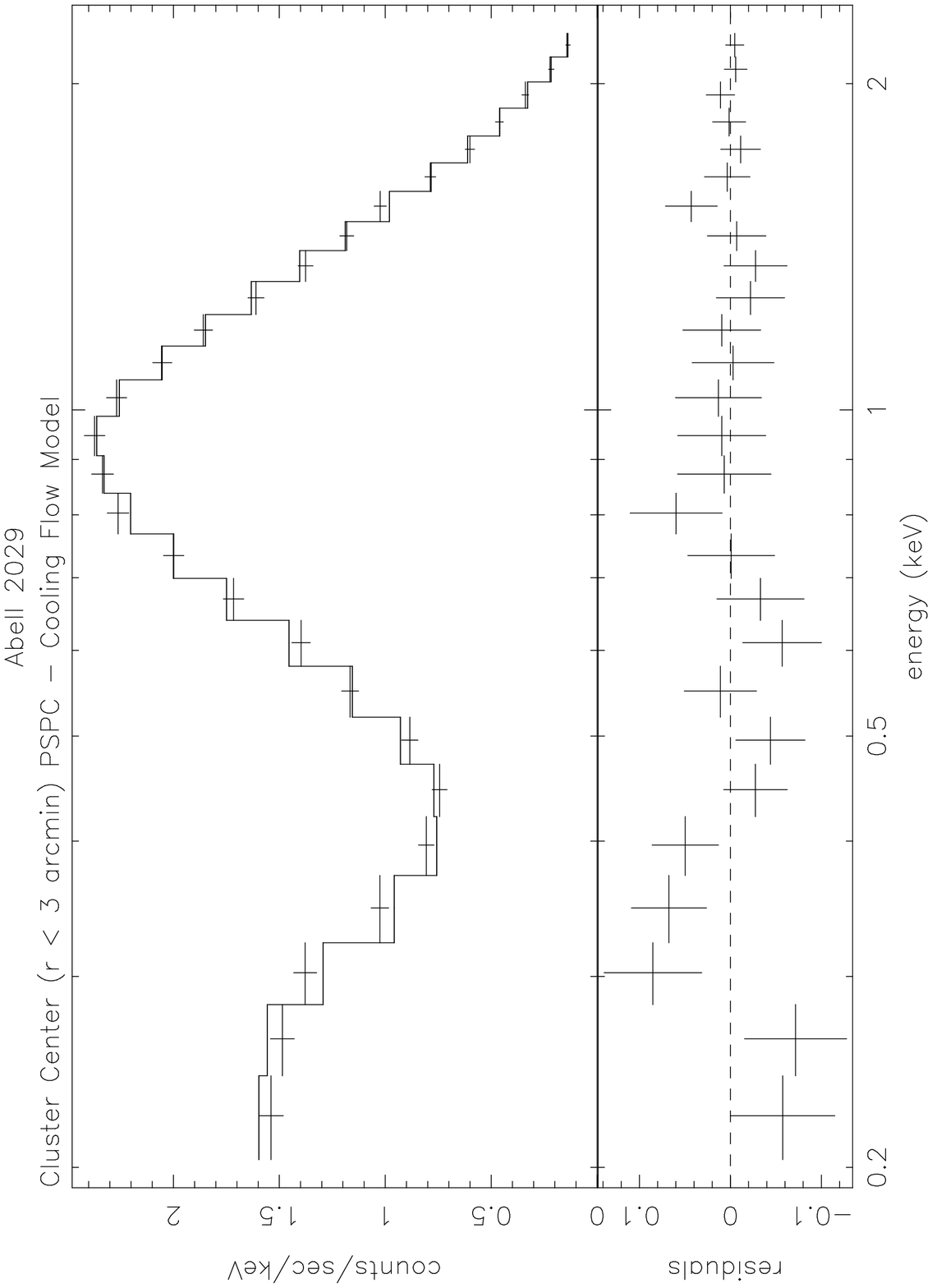}
\caption{The {\it ROSAT} PSPC X-ray spectrum for the central
cooling flow region for A2029 ($r \le 3'$).
The notation is the same as for Figure~\protect\ref{fig:center_gis_spect}.
The histogram is the best single temperature plus cooling flow
model for the PSPC spectrum
(row 20 in Table~\protect\ref{tab:spectra}).
}
\label{fig:center_pspc_spect}
\end{figure*}

Because of the inconsistent temperatures found with the different
instruments and the other evidence for a cooling flow at the
center of A2029 (\S\S~\ref{sec:intro}, \ref{sec:global},
\& \ref{sec:spatial_t}), we also fit the single temperature
plus cooling flow model to the spectra from each of the instruments.
The best-fit spectral parameters are listed in
rows 18--20 of Table~\ref{tab:spectra}.
The data, best-fit spectral models, and residuals for the three
instruments are shown in
Figures~\ref{fig:center_gis_spect},
\ref{fig:center_sis_spect}, and
\ref{fig:center_pspc_spect}.
In all three cases, the addition of a cooling flow significantly
improved the fit.
The abundances are all consistent with one another and with
the single temperature models.
The temperatures from the three instruments are now all marginally
consistent within the uncertainties.
The cooling rates are also all consistent, but the uncertainties are rather
large.
A cooling rate of about 250 $M_\odot$ yr$^{-1}$ within the central
3 arcmin would be most consistent with the three individual spectral fits.

Unfortunately, the required absorbing columns disagree even more strongly
for the cooling flow model than they did in the single temperature model.
When we tried to simultaneously fit the spectra from all three instruments
with a single model, we were unable to find any acceptable fit.
We allowed the normalization of the model to vary from instrument to
instrument to account for any error in the overall calibration of the
instruments, and only required that the model have the same spectral shape.
Still, no acceptable common fit was found.
The last row in Table~\ref{tab:spectra} shows the best-fit common
single temperature plus cooling flow model fit to all of the spectra.
It is a rather poor fit.

Examination of the residuals to such fits showed that the problem was
primarily due to a difference in the shape of the spectrum and the
fluxes at low photon energies ($\la$1 keV) between the {\it ASCA} SIS
and the {\it ROSAT} PSPC.
We tried a number of different models for the origin of the soft
X-ray absorption
(e.g., partial covering models,
intrinsic absorption models,
and models with intrinsic
absorption applied only to the cooling flow).
We also considered various alternative models for the cooling flow
spectra
(e.g., non-isobaric cooling or a finite limiting temperature for the cooling).
None of these provided an acceptable common spectral fit or a significant
improvement on the unacceptable common cooling flow fit.
These problems suggest that there are significant calibration
errors in the SIS data at low energies at present.
Similar calibration problems and overestimated absorbing columns with
the SIS have been found by other observers
(e.g., Mukai 1995).
These low energy SIS calibration uncertainties should not affect the
previous results (\S\S~\ref{sec:global} \& \ref{sec:spatial}),
because low energy SIS data was not used there.

\subsection{Excess Soft X-ray Absorption?} \label{sec:cflow_exabsorb}

White et al.\ (1991) reanalyzed the $Einstein$ Observatory Solid
State Spectrometer (SSS) spectrum of the central 3 arcmin radius
of A2029 (essentially the same region as used to extract the {\it ASCA}
and {\it ROSAT} PSPC spectra).
White et al.\ found an excess absorption column of
$\Delta N_H = 15 \times 10^{20}$ cm$^{-2}$ (if the excess absorber
is modeled as Galactic material).
In their fits to the spectrum, the absorber was treated as lying in
the foreground of all of the X-ray emission.

\begin{figure*}[tbh]
\vskip4.3truein
\includegraphics{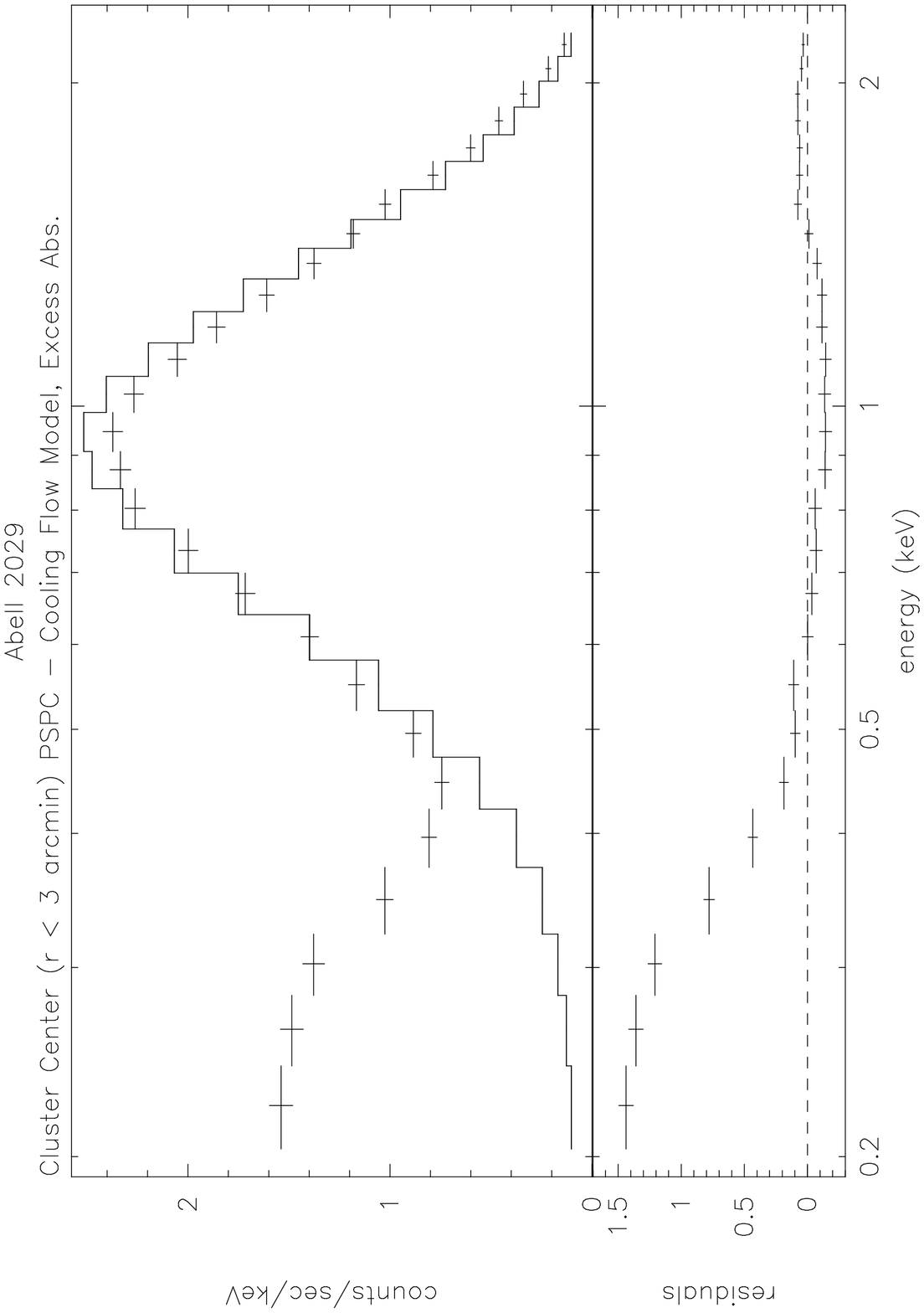}
\caption{The {\it ROSAT} PSPC X-ray spectrum for the central
3 arcmin radius of A2029 compared to the best-fit single temperature plus
cooling flow model with the absorbing column set at the value from
White et al.\ (1991).
The notation is the same as in
Figure~\protect\ref{fig:center_pspc_spect}.
}
\label{fig:center_pspc_white_spect}
\end{figure*}

Such a large foreground excess column is completely inconsistent with the
{\it ROSAT} PSPC spectrum.
The 90\% upper limit on the foreground excess absorbing column in the
inner 3 arcmin of A2029 from the PSPC spectrum is
$\Delta N_H < 2.05 \times 10^{19}$ cm$^{-2}$.
Figure~\ref{fig:center_pspc_white_spect} shows that best-fit single
temperature plus cooling flow model for the PSPC spectrum of the
inner 3 arcmin radius of A2029, with the absorption fixed at the value
from White et al.\ (1991).
This is a terrible fit to the spectrum, with a $\chi^2$ of 2417 for
23 d.o.f.
The strong soft X-ray emission in the spectrum from the center of the A2029
cluster is inconsistent with any large amount of excess foreground
absorption.
The White et al.\ value for the excess foreground absorption is also
inconsistent with the {\it ASCA} SIS and GIS spectra of the central
3 arcmin of A2029 (rows 18 and 19 in Table~\ref{tab:spectra}),
albeit at a lower level of significance.

If the excess absorption is strongly concentrated to the center of
the cooling flow, then its effects would be diluted in the 3 arcmin
aperture used to collect the cooling flow spectrum.
In order to provide a stronger constraint on excess absorption, we
also extracted the {\it ROSAT} PSPC spectrum from the inner 1 arcmin
radius of the cooling flow.
To provide a conservative upper limit on any foreground excess absorption,
we compared the absorption derived from this inner region with both
the Galactic column of $N_H = 3.07 \times 10^{20}$ cm$^{-2}$ from
Stark et al.\ (1992), and the galactic absorption derived from the
PSPC spectrum of the outer parts of the cluster (the annulus from
3--16 arcmin in radius).
We adopted the larger value of the upper limit on the absorption for
these two determinations of the Galactic column.
The excess absorber was treated as a foreground screen, but at the
redshift of the cluster (this increases slightly the limits on the
excess column).
This gave a 90\% upper limit on the column of foreground excess absorption
of
\begin{equation} \label{eq:exc_absorb_fore}
\Delta N_H (foreground) < 7.3 \times 10^{19} \, {\rm cm}^{-2} \, .
\end{equation}

The detection of excess absorption toward the central regions of
cooling flow clusters of galaxies would suggest that the absorber is
physically located with the cooling flow region
(White et al.\ 1991;
Allen et al.\ 1993;
Irwin \& Sarazin 1995).
In this case, it is much more difficult to detect or constrain
the excess absorption.
To illustrate this, we consider a model in which the excess
absorption is applied only to the cooling flow component of the spectrum,
but not to the ambient cluster emission.
If this model is fit to the PSPC spectra of the inner 3 arcmin of A2029,
the best-fit value of the excess absorption is still zero.
However, the 90\% upper limit on the excess absorption is very large,
\begin{equation} \label{eq:exc_absorb_cf}
\Delta N_H (cooling~flow) < 2.6 \times 10^{21} \, {\rm cm}^{-2} \, .
\end{equation}
The reason for this very weak limit is that the value of the excess
absorption and the cooling rate are very strongly correlated in the
fits.
If one increases the amount of soft X-ray emission by increasing the
cooling rate $\dot{M}_{cool}$, one can increase the excess soft X-ray
absorption $\Delta N_H $ if it is applied only to this cooling flow
emission.

Our conclusion is that the PSPC spectra (and, to a lesser extent, the
{\it ASCA} spectra) rule out the large foreground absorption claimed
by White et al.\ (1991).
However, an absorber confined to the cooling flow region is much more
difficult to detect or constrain with PSPC spectra.
In principle, {\it ASCA} SIS spectra could give a stronger constraint,
but the inconsistency between the {\it ASCA} SIS and {\it ROSAT} PSPC
spectra in A2029 suggests that there are still significant calibration
uncertainties with the {\it ASCA} SIS spectra at low energies.

For the purposes of determining the X-ray gas pressure in the inner region
where the radio source is located (Huang \& Sarazin 1998), we also
determined the best-fit PSPC temperature for the inner arcmin.
This gave $kT = 3.16^{+0.44}_{-0.35}$ keV.

\section{MASS PROFILE} \label{sec:masses}

\begin{figure*}[tbh]
\vskip3.75truein
\includegraphics{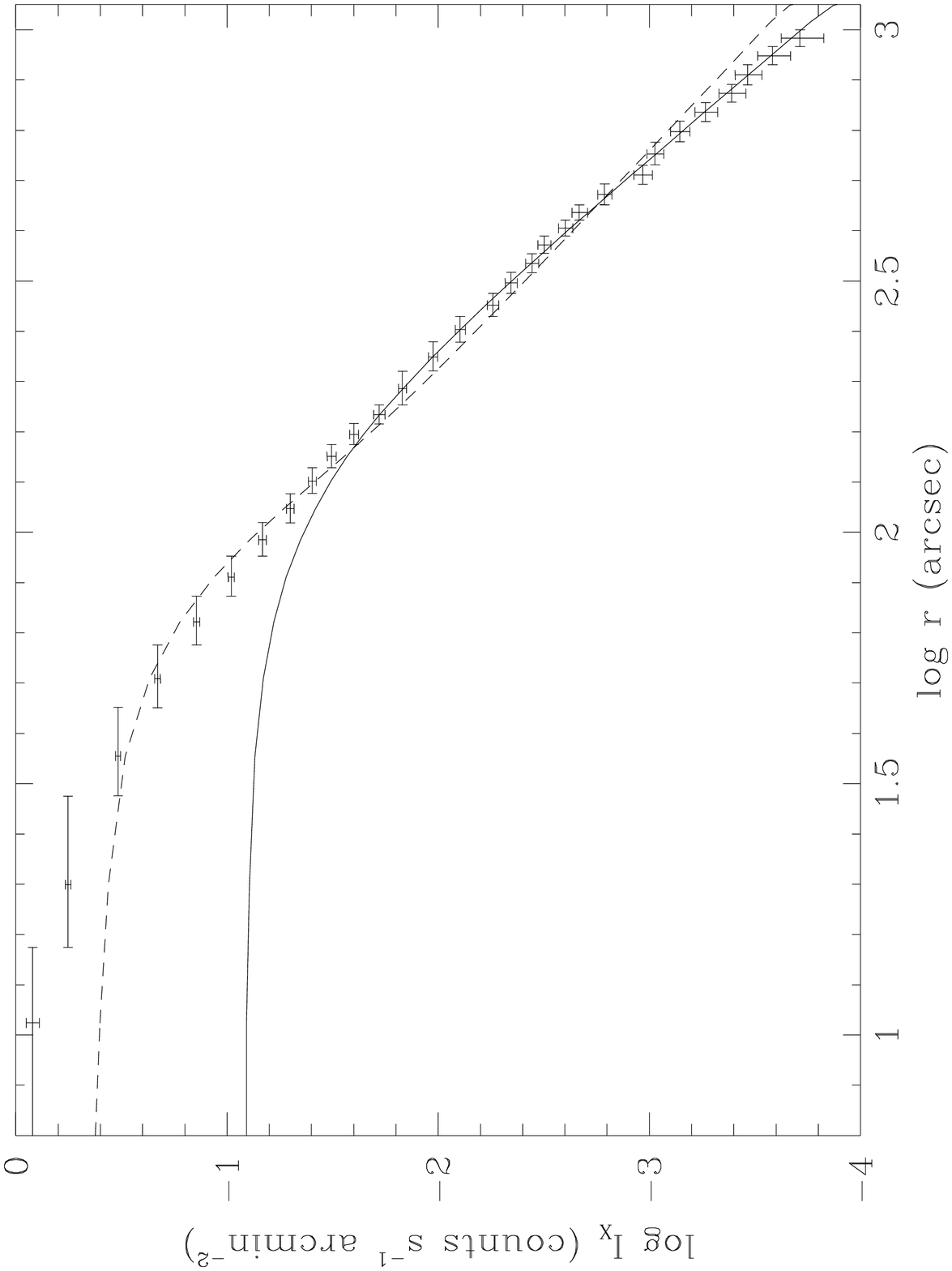}
\caption{The {\it ROSAT} PSPC X-ray surface brightness profile
for the spectral band 0.5--2.0 keV.
The horizontal error bars show the width of the annulus used to collect
the counts.
The vertical error bars give the 90\% confidence region.
The surface brightness is plotted at the emission-weighted median radius
for that annulus.
The dashed curve is the unacceptable beta-model fit to the surface
brightness including all of the data points.
The solid curve is the acceptable beta-model fit, excluding the cooling
flow regions ($r < 164''$).
}
\label{fig:pspc_profile}
\end{figure*}

We used the information on the temperature and temperature distribution
from the {\it ASCA} spectra to determine the gas and total mass
distribution in A2029.
The first step in this process was to determine the X-ray surface
brightness distribution, using the {\it ROSAT} PSPC image
(Fig.~\ref{fig:pspc_image}).
The X-ray count rate was determined in concentric circular annuli,
excluding regions around each of the point sources in the field.
The surface brightness was corrected for background using the
same regions used for the background correction of the spectra
(\S~\ref{sec:pspc_data}).
The resulting X-ray surface brightness profile for the spectral
band 0.5--2.0 keV is shown in Figure~\ref{fig:pspc_profile}.
A sufficient number of counts were collected in each annulus to insure
that gaussian statistics applied.
Possible faint emission is seen extending to a radius of at least 22
arcmin ($2.64 h_{50}^{-1}$ Mpc),
but the uncertainties in the surface brightness are very
large, due to the low surface brightness and the effect of the
instrumental support structure in this region.
In the subsequent analysis (and in Fig.~\ref{fig:pspc_profile}),
we consider only the surface brightness at radii $r < 18$ arcmin
($2.16 h_{50}^{-1}$ Mpc) where the uncertainties are more reasonable. 

We fitted the X-ray surface brightness in Figure~\ref{fig:pspc_profile}
with the isothermal ``beta'' model
\begin{equation}
I_X (r) = I_o \left[ 1 + \left( \frac{r}{r_{core}} \right)^2
\right]^{-3 \beta + 1/2} \, .
\label{eq:beta_model}
\end{equation}
The model was convolved with the PSPC PSF and compared to the
data.
When all of the data points in Figure~\ref{fig:pspc_profile} were
included, no acceptable fit could be found.
The best-fit model, shown as a dashed curve in Figure~\ref{fig:pspc_profile},
had $\chi^2 = 1074.5$ for 27 d.o.f.
As has generally been found with cooling flows clusters
(e.g., Jones \& Forman 1984),
the beta model does not fit the sharply peaked central surface
brightness within the cooling radius.
Thus, we fit the surface brightness with the beta model, progressively
removing the interior points until an acceptable fit
($\chi^2/$d.o.f $< 1.4$) was found.
This required that the data interior to 164$''$ ($0.33 h_{50}^{-1}$ Mpc)
be eliminated.
When these points were removed, an acceptable fit with
$r_{core} = 108^{+22}_{-16}$ arcsec and
$\beta = 0.64 \pm 0.02$ was found.
Our value of $\beta$ is in very good agreement with that given by
David et al.\ (1995) of $\beta = 0.65^{+0.03}_{-0.02}$.
Our fit, which gave
$\chi^2 = 18.9$ for 16 d.o.f.\ ($\chi^2/$d.o.f $= 1.182$), is shown as the
solid curve in Figure~\ref{fig:pspc_profile}.
Because the derived core radius is within the region of data excluded from
the fit and where the beta-model doesn't fit, the value of the core radius
and central surface brightness $I_o$ are probably not given reliably by
this fit.
However, this model does appear to provide a reasonable fit for the
surface brightness beyond about 160 arcsec.
Because we are most interested in the mass at large radii, we will use this
fit to describe the gas distribution.

We adopt the linear fit to the radial temperature in the gas derived from
the {\it ASCA} spectra (eq.~\ref{eq:t_linear}).
Assuming the abundances of the heavy elements are 0.40 of the solar value,
which was the best-fit abundance for the global cluster spectrum,
we inverted the best-fit beta-model to determine the density
of the gas $\rho_{gas}$ and its pressure $P$.
The density of the gas was integrated over the interior volume to 
give the interior gas mass as a function of the radius, $M_{gas} (r)$.
The total gravitational mass was determined from the
assumption that the gas is in hydrostatic equilibrium, which implies that
\begin{equation}
M_{tot} (r) = - \frac{r^2}{G \rho_{gas}} \, \frac{d P}{d r} \, .
\label{eq:hydrostatic}
\end{equation}
The uncertainties in $M_{gas} (r)$ and $M_{tot} (r)$ were determined from the
uncertainties and covariances in the coefficients of the fit to the temperature
as a function of radius
(eq.~\ref{eq:t_linear} and Fig.~\ref{fig:asca_temp})
and in the PSPC surface brightness fit
(eq.~\ref{eq:beta_model} and Fig.~\ref{fig:pspc_profile}).

\begin{figure*}[tbh]
\vskip5.0truein
\includegraphics{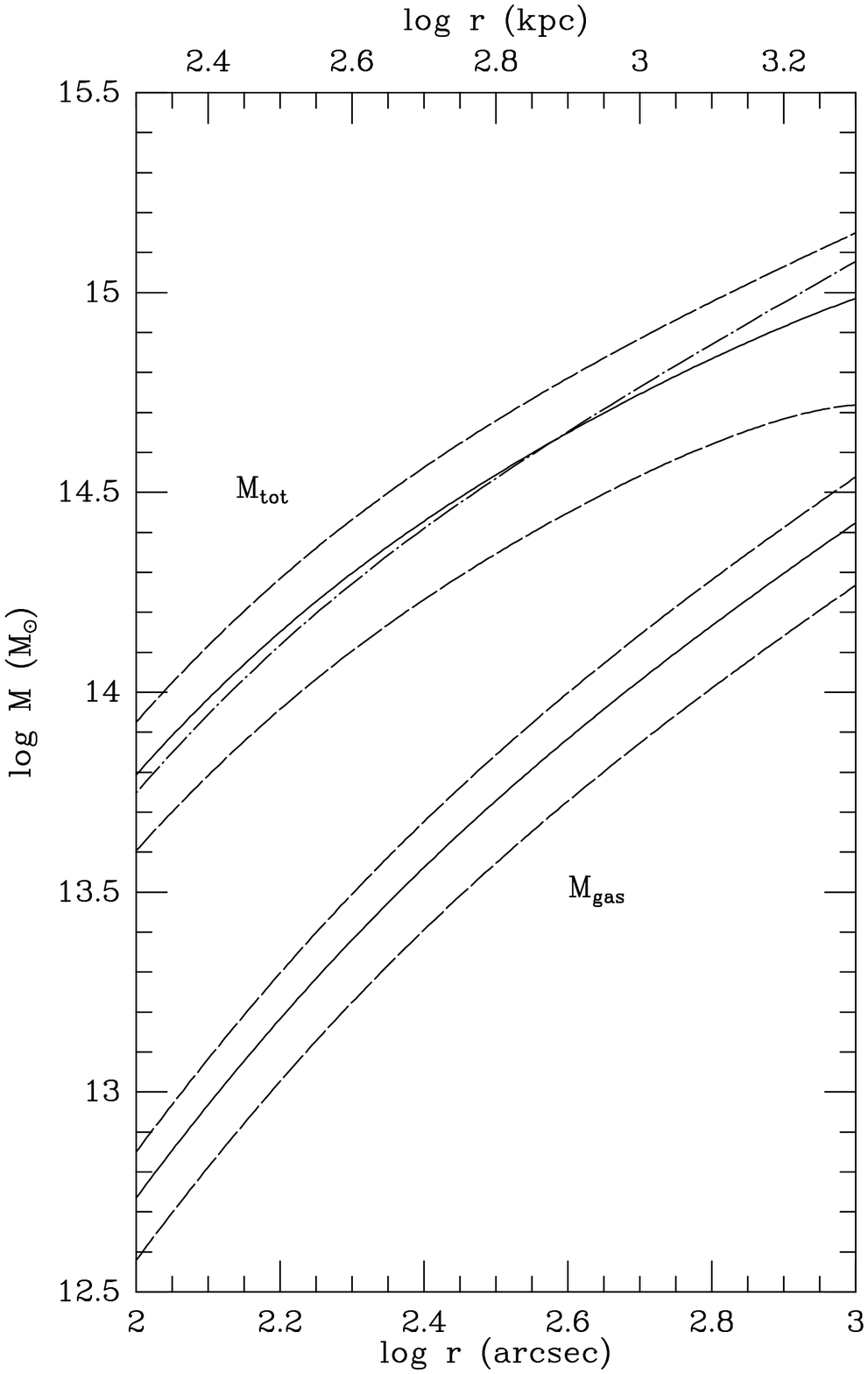}
\caption{The total gravitating mass and gas mass as a function
of radius determined from the {\it ASCA} spectra and {\it ROSAT} PSPC surface
brightness profile of A2029.
The upper and lower solid curves give the total mass and gas mass,
respectively.
The dashed curves give the 90\% confidence interval for each of these
quantities.
The dash-dot curve near the upper solid curve gives the total mass if
the gas is assumed to be isothermal at the best-fit single temperature,
rather than having a linear temperature variation with radius.
The radial scale is given in arcsec at the bottom of the plot, and in
kpc at the top.
All values assume $H_o = 50 h_{50}$ km s$^{-1}$ Mpc$^{-1}$.
The masses may be inaccurate at the smallest radii because of the effect
of the cooling flow.}
\label{fig:masses}
\end{figure*}

The resulting mass profiles are shown in Figure~\ref{fig:masses}.
At a radius of 16$'$ ($1.92 h_{50}^{-1}$ Mpc, the largest region used for
the temperature determination), the masses are
$M_{gas} = ( 2.52 \pm 0.77 ) \times 10^{14} h_{50}^{-5/2} \, M_\odot$ and
$M_{tot} = ( 9.42 \pm 4.22 ) \times 10^{14} h_{50}^{-1} \, M_\odot$,
and the ratio of masses is
$M_{gas}/M_{tot} = ( 0.26 \pm 0.14 ) h_{50}^{-3/2}$.
The fraction of the total cluster mass which is due to the
intracluster gas is rising steeply with radius at the largest radii.
A very similar result was found for A2029 by
David et al.\ (1995) assuming an isothermal gas profile, and for
Abell~2256 by Markevitch \& Vikhlinin (1997) using an {\it ASCA} temperature
profile.

\section{CONCLUSIONS} \label{sec:conclude}

We have analyzed the {\it ASCA} and {\it ROSAT} PSPC X-ray observations
of the rich cluster Abell 2029.
The {\it ASCA} GIS and {\it ROSAT} PSPC are in good agreement on the
global X-ray spectrum, and give an average ambient gas temperature of
$9.35^{+0.55}_{-0.45}$ keV (including the effects of a cooling flow).
If the gas temperature is assumed to be constant in the analysis of
the spatially resolved {\it ASCA} SIS and GIS, the average temperature is
found to be 8.6 keV.
The iron abundance in the gas is $0.40 \pm 0.04$ of the solar value.
There is no significant evidence for any variation in the abundance
with position in the cluster.

The global X-ray spectra, central X-ray spectra, and {\it ROSAT}
surface brightness all require a cooling flow at the cluster center.
The global X-ray spectrum implies that the total cooling rate is
$363^{+79}_{-96} h_{50}^{-2} \, M_\odot$ yr$^{-1}$.

The global X-ray spectra are consistent with the Galactic
value for the soft X-ray absorption toward the cluster.
The {\it ROSAT} PSPC spectra of the central regions of the
cluster of completely inconsistent with the large value
of foreground excess absorption found by White et al.\ (1991)
based on the {\it Einstein} SSS spectrum.
The upper limit on excess foreground absorption is
$7.3 \times 10^{19}$ cm$^{-2}$.
However, the spectra do not rule out a significant amount of intrinsic
absorbing gas located within the cooling flow region.
One concern with modeling the cooling flow spectrum and absorption
in A2029 is that the {\it ASCA} SIS and {\it ROSAT} PSPC spectra do
not appear to be consistent.
This may indicate that there are calibration problems with the
{\it ASCA} SIS spectra at low energies.

The {\it ASCA} spectra of the cluster indicate that the gas
temperature declines with radius.
Similar declines are seen in most clusters studied with
{\it ASCA}
(Markevitch et al.\ 1997).

The PSPC image shows that the cluster is elliptical, but is very regular
and smooth.
This agrees with previous analyses of the optical and X-ray distribution
of the cluster
(Slezak et al.\ 1994;
David et al.\ 1995;
Buote \& Tsai 1996).
We also find that there is no significant evidence for any irregularities
in the temperature distribution in the cluster, as would be produced by a
subcluster merger.
Structure in the X-ray surface brightness and particularly the gas
temperature distribution has been interpreted as evidence for mergers
and other hydrodynamic activity in many other clusters
(e.g., Henry \& Briel 1995;
Markevitch et al.\ 1996,1997).
The lack of structure in the X-ray properties of A2029 suggests that
the cluster is relaxed and that the gas is in hydrostatic equilibrium.

We use the assumption of hydrostatic equilibrium to determine the
gravitational mass of the cluster as a function of radius.
Within a spherical radius of $16'$ ($1.92 h_{50}^{-1}$ Mpc),
the total gravitational mass is
$M_{tot} = ( 9.42 \pm 4.22 ) \times 10^{14} h_{50}^{-1} \, M_\odot$,
while the mass of gas is
$M_{gas} = ( 2.52 \pm 0.77 ) \times 10^{14} h_{50}^{-5/2} \, M_\odot$.
The gas fraction of the cluster at this radius is
$M_{gas}/M_{tot} = ( 0.26 \pm 0.14 ) h_{50}^{-3/2}$,
and the gas fraction is is increasing with radius at the largest radii.
Thus, A2029 is a particularly strong example of the so-called
``baryon catastrophe'' in clusters
(e.g., David et al.\ 1995).

\acknowledgements

We thank the referee, Joel Bregman, for a number of helpful suggestions.
C. L. S. was supported in part by NASA ROSAT grants NAG 5--1891,
NAG 5--3308, NASA ASCA grant NAG 5-2526, and NASA Astrophysical
Theory Program grant 5-3057.
M. L. M. was supported by NASA grant NAG5-2611.


\begin{references}

\reference{}
Abell, G. O., Corwin, H. G., \& Olowin, R. P. 1989,
ApJS, 70, 1

\reference{}
Allen, S. W., Fabian, A. C., Johnstone, R. M., White, D. A.,
Daines, S. J., Edge, A. C., \& Stewart, G. C. 1993,
MNRAS, 262, 901

\reference{}
Anders, E., \& Grevesse, N. 1989, Geochim.\ Cosmoshim.\ Acta, 53, 197

\reference{}
Arnaud, K. A. 1989, preprint

\reference{}
Buote, D. A., \& Tsai, J. C. 1996, ApJ, 458, 27

\reference{}
David, L., Jones, C., \& Forman, W. 1995, ApJ, 445, 578

\reference{}
David, L. P., Slyz, A., Jones, C., Forman, W., Vrtilek, S. D., \& Arnaud,
K. A. 1993, ApJ, 412, 479

\reference{}
Henry, J. P, \& Briel, U. G. 1995, ApJ, 443, L9

\reference{}
Huang, Z., \& Sarazin, C. L. 1996, ApJ, 461, 622

\reference{}
Huang, Z., \& Sarazin, C. L. 1998, ApJ, 496, in press

\reference{}
Irwin, J. A., \& Sarazin, C. L. 1995, ApJ, 455, 497

\reference{}
Jones, C. \& Forman, W. 1984, ApJ, 276, 38

\reference{}
Kaastra, J. S., Mewe, R., Liedahl, D. A., Singh, K. P., White, N. E., \&
Drake, S. A. 1996, A\&A, 314, 547

\reference{}
Liedahl, D. A., Osterheld, A. L., \& Goldstein, W. H. 1995, ApJ, 438, L115

\reference{}
Markevitch, M., Mushotzky, R., Inoue, H., Yamashita, K., Fukuzawa, A., \&
Tawara, Y. 1996, ApJ, 456, 437

\reference{}
Markevitch, M. L., Forman, W., Sarazin, C. L., \& Vikhlinin, A. 1997,
ApJ, submitted

\reference{}
Markevitch, M. L., \& Vikhlinin, A. 1997, ApJ, in press

\reference{}
McNamara, B. R., \& O'Connell, R. W. 1989, AJ, 98, 2018

\reference{}
Mukai, K. 1995, Minutes of the ASCA Calibration Workshop \#2,
http://heasarc.gsfc.nasa.gov/docs/asca/cal\_probs.html

\reference{}
Plucinsky, P. P., Snowden, S. L., Briel, U. G., Hasinger, G.,
\& Pfeffermann, E. 1993, ApJ, 418, 519

\reference{}
Press, W. H., Teukolsky, S. A., Vetterling, W. T., \& Flannery, B. P. 1992,
Numerical Recipes (Cambridge: Cambridge Univ. Press)

\reference{}
Raymond, J. C., \& Smith, B. W. 1977,
ApJS, 35, 419

\reference{}
Sarazin, C. L., O'Connell, R. W., \& McNamara, B. R. 1992,
ApJ, 389, L59

\reference{}
Slezak, E., Durret, F., \& Gerbal, D. 1994, AJ, 108, 1996

\reference{}
Snowden, S. L. 1995, Cookbook for Analysis Procedures for ROSAT
XRT/PSPC Observations of Extended Objects and the Diffuse
Background (Greenbelt: NASA USRSDC)

\reference{}
Stark, T., et al.\ 1992, ApJS, 79, 77

\reference{}
Takahashi, T., Markevitch, M., Fukazawa, Y., Ikebe, Y., Ishisaki, Y.,
Kikuchi, K., Makishima, K., \& Tawara, Y. 1995, ASCA Newsletter, No.\ 3

\reference{}
White, D. A., Fabian, A. C., Johnstone, R. M., Mushotzky, R. F., \&
Arnaud, K. A. 1991, MNRAS, 252, 72

\reference{}
Wise, M. W., \& Sarazin, C. L. 1993, ApJ, 415, 58

\end{references}
\end{document}